\documentclass[11pt,a4paper]{article}
\pdfoutput=1

\usepackage{amsmath}
\usepackage{amsfonts}
\usepackage{amsthm}
\usepackage{amssymb}
\usepackage{amscd}
\usepackage[british]{babel}
\usepackage{graphicx}
\usepackage{psfrag}
\usepackage{epsfig}
\usepackage{rotating}
\usepackage{times}

\theoremstyle{plain}

\newtheorem{remark}{Remark}[section]

\newcommand{\boxend}{\flushright{$\Box$}}

\newcommand{\f}{\frac}

\renewcommand{\tilde}{\widetilde}

\begin{document}

\title{
Viability of the matter bounce scenario in $F(T)$ gravity and Loop Quantum Cosmology for general potentials}

\author{Jaume Haro$^{a,}$\footnote{E-mail: jaime.haro@upc.edu}  and Jaume Amor\'os$^{a,}$\footnote{E-mail: jaume.amoros@upc.edu}
}

\maketitle

{$^a$Departament de Matem\`atica Aplicada I, Universitat
Polit\`ecnica de Catalunya, Diagonal 647, 08028 Barcelona, Spain \\
}

\thispagestyle{empty}

\begin{abstract}
We consider the matter bounce scenario in $F(T)$ gravity and Loop Quantum Cosmology (LQC) for phenomenological potentials that at early times
provide a nearly matter dominated Universe in the contracting phase,  having a
reheating mechanism in the expanding or contracting  phase, i.e., being able to release the energy of the scalar field creating  particles that thermalize in order to match
with
the hot Friedmann Universe, and finally at late times leading to the current cosmic acceleration.
For these potentials, numerically solving the dynamical perturbation equations we have seen that,
for the particular $F(T)$ model  that we will name
{\it teleparallel version of LQC}, and whose modified Friedmann equation  coincides with the corresponding one in  holonomy corrected LQC when one deals with the flat 
Friedmann-Lema{\^\i}tre-Robertson-Walker (FLRW) geometry, 
the corresponding  equations obtained from the well-know  perturbed equations in $F(T)$ gravity
lead to theoretical results that fit well with current observational data.
More precisely, in this {\it teleparallel version of  LQC} there is a
set of solutions which leads to theoretical results that match correctly with last BICEP2 data, and there is another set whose theoretical results fit well with {\it Planck's}
experimental data. On the other hand, in the standard holonomy corrected LQC, using the perturbed equations obtained replacing the Ashtekar connection by a suitable sinus function and
inserting some counter-terms in order to preserve the algebra of constrains, the  theoretical value of the tensor/scalar ratio is smaller than
in the teleparallel version, which means that
there is always a set of solutions that matches with {\it Planck's} data, but for some potentials BICEP2 experimental results disfavours holonomy corrected LQC.
\end{abstract}

\vspace{0.5cm}

{\bf Pacs numbers:}  04.80.Cc, 98.80.Bp, 98.80.Qc, 04.60.Pp, 04.50.Kd


\section{Introduction}
It's well-known that inflation suffers from several problems (see \cite{brandenberger} for a review about these problems), like
 the initial singularity which normally  is not addressed (as an exception, in \cite{bv} the problem was addressed concluding that the initial singularity is unavoidable in general
  inflationary models), the fine-tuning of the degree of flatness required for the potential in order to achive successful inflation \cite{ffg91},  
  or the following problem  related with  initial conditions:
In inflationary cosmology it is usually assumed  that modes well inside the Hubble radius are initially (at the beginning of inflation) in the adiabatic vacuum in order to obtain a nearly scale invariant spectrum. This
 assumption could be accepted if, as in Linde's early papers about chaotic inflation (see for instance \cite{l84} for a review), inflation started at energy densities of the
 order of Planck's scale, because in that case before
inflation it would be impossible to describe classically our Universe. However, from
the four-year data set provided by the Cosmic Background Explorer (COBE) satellite or from the seven-year data of Wilkinson Microwave Anisotropy Probe (WMAP), we know that
the observed value of the power spectrum for scalar perturbations is constrained to be
$P_{\zeta}(k)\cong 2\times 10^{-9}$ \cite{btw06} for modes that exit the Hubble radius $60$ e-folds before the end of inflation, which means  that in chaotic inflation the slow roll phase
started at energy densities of the
order $10^{-11}\rho_{pl}$, and consequently, the evolution of the Universe could be described classically before inflation. Then, it is essential to
know the evolution of the modes before inflation, because if they re-enter in the Hubble radius, positive and negative frequencies could mix, and thus,
those modes would not be in the vacuum state.

In order to avoid these problems,
an alternative scenario to the inflationary paradigm, called {\it matter bounce scenario} \cite{np08} (essentially it depicts, at very early times, a matter dominated Universe in the
contracting phase that evolves towards the bounce to enter in the expanding phase), has been developed to
explain the evolution of our Universe. This model, like inflation, solves the horizon problem that appears in Einstein Cosmology (EC) and improves the flatness problem
in EC (where spatial flatness is an unstable fixed point and  fine tuning of initial conditions is required),
because
the contribution of the spatial curvature decreases in
the contracting phase at the same rate as it increases in the expanding one (see for instance \cite{brandenberger1}). However, it suffers from the
anisotropy problem that does not exist in other models like Ekpyrotic scenarios \cite{ewst}, for this reason
an improved model combining  both scenarios could avoid this problem \cite{cbp}, and become a realistic alternative to inflation.

There are essentially  two ways to set up a
matter bounce scenario in the flat FLRW geometry: within the framework of EC violating the null energy condition at the bouncing point
\cite{cqplz07}, or
going beyond EC.
In order to  violate the null energy condition in EC one needs to incorporate
new forms of matter such as phantom \cite{aw}
or quintom fields \cite{cqplz07}, Galileons \cite{qeclz} or phantom condensates \cite{lbp}.
Going beyond EC one can introduce higher derivatives in the action \cite{mb}, braneworld bouncing scenarios \cite{ss}, Ekpyrotic \cite{kost},
Pre-Big-Bang \cite{gv}, loop quantum \cite{svv} or teleparallel cosmologies \cite{aho13} ,
modified $f(R)$ gravity \cite{bmmno}, extended  loop quantum cosmology to $R^2$ gravity \cite{aho14},
etc.

In the present work  we only  deal with the matter bounce scenario,
that in the flat FLRW geometry has the same Friedmann equation as in  holonomy corrected  Loop Quantum Cosmology (LQC), which leads to the simplest bouncing scenario, and 
where the numerical calculation can be carried out completely.

We will perform a deep and detailed study of the
 evolution of cosmological perturbations in this scenario, that improves on those recently  made in \cite{w13}  using standard holonomy corrections and those of
\cite{h13}  
where an exemple of $F(T)$ gravity which leads, in the flat FLRW geometry, to the same modified Friedmann equation as in holonomy corrected LQC, was used to obtain
the perturbation equations in the framework of $F(T)$ gravity. In both works,
the potential of the used scalar field  leads to solutions that at early and late times are in a matter dominated phase, and do not agree with the current acceleration of the Universe.
Moreover, since only one analytic solution of the isotropic equations (the unperturbed ones) is known, all the calculations are performed with this analytic solution,
which leads to conclusions that do not agree with the current observations.
For example, in those works it is claimed that, in order to match theoretical results with observations, the value of the critical density in LQC has to be of the order of $10^{-9}\rho_{pl}$
which contradicts the current value $0.4\rho_{pl}$  \cite{meissner}. Dealing with tensor perturbations, in holonomy corrected LQC the equation of tensor perturbations has singularities when
the energy density is a half of the critical one, meaning that one could consider infinitely many mode solutions
because there is not a criterium of continuity at the singular point to decide which mode is the correct one, and thus, there is not a unique way to calculate the power spectrum for
tensor perturbations.
On the other hand, in the {\it teleparallel version of  LQC}, i.e. using the  $F(T)$-perturbed equation for the model whose isotropic Friedmann equation coincides with the 
holonomy corrected one of LQC, there is a unique way to calculate this power spectrum, but for the analytical solution of the isotropic equations the ratio of tensor to
scalar perturbations is approximately equal to $6$, which does not agree with the current observational data.

However, for the other solutions which we have obtained numerically in this work, there is a a set of solutions that fit well with the recent BICEP2 data
\cite{Kovac}, i.e. for the solutions that belong in that
set its tensor/scalar ratio satisfy  $r=0.20^{+0.07}_{-0.05}$ \cite{ha14}, and another one whose tensor/scalar ratio is smaller than $0.11$, matching
correctly with the latest {\it Planck's} data \cite{Ade}. Then,
our main objective in this work is to generalize this result to more phenomenological potentials (containing a matter domination in the contracting phase at early times, and leading to a reheating
process in the expanding
phase in order to match with the current $\Lambda$CDM model of the Universe, or more generally, with the hot Friedmann Universe plus
 the current cosmic acceleration). That is, we want  to find phenomenological potentials that have a  set of solutions that agrees
 with BICEP2 data, and another one that matches correctly with the latest {\it Planck's} results. For some of those potentials we present numerical results supporting this match.

\vspace{0.25cm}
The outline of the paper goes as follows: In Section II we review the way to obtain the Mukhanov-Sasaki equations in standard holonomy corrected LQC
and in its teleparallel version, i.e. in $F(T)$ gravity for the  model that, restricted to the flat FLRW geometry, leads to the same dynamical equations that in holonomy corrected LQC.  Section III is devoted  to showing that,
for a matter bounce scenario, the relation
between the
Bardeen potential and the curvature fluctuation in co-moving coordinates, in Fourier space, when the Universe is matter dominated at late times,
is the same as in inflationary cosmology when the Universe has a phase transition from the quasi de Sitter stage to the matter dominated one, that is,
its quotient is equal to $3/5$. As a consequence, since in
 matter bounce scenario the
power spectrum of the curvature fluctuation in co-moving coordinates is scale invariant, the corresponding power spectrum for the Bardeen potential is also scale invariant, which is not
trivial to show in this scenario. Section IV is destined to review, with all the details, the calculation of the power spectrum of scalar perturbations in both
holonomy corrected LQC and its teleparallel version. In Section V we deal with the problems of the potential
currently used to study the matter bounce scenario in holonomy corrected LQC, for example the
absence of an explanation for the current cosmic acceleration. In the last Section we suggest some models that
include a reheating process and the current acceleration of the Universe, and which
lead to solutions whose
theoretical results match correctly with current observational data (power spectrum of scalar perturbations, spectral index of scalar perturbations and
ratio of tensor to scalar perturbations). Moreover, at the end of this Section we perform a detailed study of the reheating
in the matter bounce scenario via gravitational particle production.

\vspace{0.25cm}

The units used in the work are $\hbar=c=8\pi G=1$.

\section{Mukhanov-Sasaki variables in LQC}
Assuming that the dynamics of the Universe is carried on by a scalar field, namely $\varphi=\bar{\varphi}+\delta\varphi$, where $\bar{\varphi}$ is
the homogeneous part of the field,
in EC, where in the flat Friedmann-Lema{\^\i}tre-Robertson-Walker (FLRW) geometry the Friedmann equation is $H^2=\frac{\rho}{3}$,
the perturbation equations, in the longitudinal gauge, are (see for instance \cite{m05})
\begin{eqnarray}
 \frac{1}{a^2}\Delta\Phi=\frac{\dot{\bar{\varphi}}^2}{2H}\frac{d}{dt}\left(\frac{H}{\dot{\bar{\varphi}}}\delta\varphi+\Phi \right),\quad
\frac{d}{dt}\left(\frac{a\Phi}{H}\right)=\frac{a\dot{\bar{\varphi}}^2}{2H^2}\left(\frac{H}{\dot{\bar{\varphi}}}\delta\varphi+\Phi \right).
\end{eqnarray}

Introducing the variables
\begin{eqnarray}
 v=a(\delta\varphi+\frac{\dot{\bar{\varphi}}}{{H}}\Phi); \quad z=\frac{a\dot{\bar{\varphi}}}{{H}},\quad u=\frac{2\Phi}{\dot{\bar{\varphi}}};
\quad\theta=\frac{1}{z},
\end{eqnarray}
one obtains the  Mukhanov-Sasaki (M-S) equations
\begin{eqnarray}\label{1}
 {c}_{s}\Delta u= z\left(\frac{v}{z}\right)';\quad \theta\left(\frac{u}{\theta}\right)'= {c}_{s}v,
\end{eqnarray}
where in EC, the velocity of sound, namely $c_s$, is equal to $1$.

On the other hand, in holonomy corrected  Loop Quantum Cosmology (LQC), in the flat FLRW geometry the corresponding
modified Friedmann equation is (see for a review \cite{as11} of LQC)
\begin{eqnarray}\label{friedmann}
H^2=\frac{\rho}{3}\left(1-\frac{\rho}{\rho_c}\right),\end{eqnarray}
where $\rho_c$ is the so-called ``critical density'' (the energy at which the Universe bounces).

\begin{remark}
 It is important to realize that this equation could be obtained for all kind of fluids and scalar fields from the 
 holonomy corrected Hamiltonian in Loop Quantum Gravity (LQG), which for the flat FLRW geometry, and working in the so-called {\it new quantization scheme} \cite{corichi}, has the form
 \cite{abl03,t01,aps06a}
\begin{eqnarray}
{\mathcal H}_{LQC}\equiv-\frac{2}{\gamma^3\lambda^3}
\sum_{i,j,k}\varepsilon^{ijk} Tr[
h_i(\lambda)h_j(\lambda)h_i^{-1} (\lambda)
h_j^{-1}(\lambda)h_k(\lambda)
\{h_k^{-1}(\lambda),V\}]+\rho V,
\end{eqnarray}
where $V=a^3$ is the volume, $\gamma$ is the Barbero-Immirzi parameter (whose value is constrained, but not fixed 
as earlier believed,
by the Bekenstein-Hawking formula for the entropy of a black hole (see \cite{meissner, dreyer} for the earlier determination and \cite{perez} for updated derivation which shows
that the Immirzi parameter is no longer fixed, but only bounded in the LQC setting, by this formula)
and $\lambda=\sqrt{\frac{\sqrt{3}}{4}\gamma}$ is the  square root of the minimum eigenvalue of the
area operator in LQG.

The holonomies are given by
\begin{eqnarray}
h_j(\lambda)\equiv e^{-i\frac{\lambda \beta}{2}\sigma_j} =\cos\left(\frac{\lambda
\beta}{2}\right)-i\sigma_j\sin\left(\frac{\lambda \beta}{2}\right),
\end{eqnarray}
where  the Pauli matrices $\sigma_j$  have been used and $\beta$ is  canonically conjugate to $V$, with Poisson bracket  $\{V,\beta\}=\frac{\gamma}{2}$.

A simple calculation
leads to
the following holonomy corrected Hamiltonian 
\begin{equation}\label{3}
 {\mathcal H}_{LQC}=-3V\frac{\sin^2(\lambda\beta)}{\lambda^2\gamma^2 }+\rho V.
\end{equation}

Then, from the Hamilton equation $\dot{V}=\{V,{\mathcal H}_{LQC}\}$ one obtains the relation
 $H=\frac{\sin(2\lambda\beta)}{2\lambda\gamma }$,
that together with the Hamiltonian constrain ${\mathcal H}_{LQC}=0$, lead to the
  holonomy corrected Friedmann equation in LQC, which
depicts, in the plane $(H,\rho)$,
 the ellipse given by (\ref{friedmann}). Note also that in formula (\ref{friedmann}) 
 the critical density is equal to 
 \begin{eqnarray}\label{criticaldensity}
 \rho_c=\frac{3}{\lambda^2\gamma^2}=\frac{4\sqrt{3}}{\gamma^3}.
 \end{eqnarray}
 
 This method to obtain the holonomy corrected Friedmann equations was obtained independently in \cite{he,b,dmp}, and does not depend  of the fluid or scalar field used to depict the
 material composition of the Universe. Another different question is how to obtain equation (\ref{friedmann}) from the quantum Hamiltonian constrain in LQC. 
 In this situation, using coherent states, it  has been
 proved  that equation (\ref{friedmann}) is obtained for models without potential \cite{aps06}, and what has not been confirmed yet, is whether this equation (\ref{friedmann}) could
 be recovered from the quantum Hamiltonian constrain in all kind of fluids and scalar fields.
\end{remark}

In holonomy corrected LQC
the perturbation equations were obtained for the first time in \cite{cmbg12} using holonomy corrections, i.e.,
replacing the Ashtekar connection by a suitable sinus function \cite{bh08}, and adding some counter-terms to the
perturbed Hamiltonian in order to preserve the algebra of constrains.
These equations are
\begin{eqnarray}
 \frac{1}{a^2}\Delta\Phi=\frac{\dot{\bar{\varphi}}^2}{2H}\frac{d}{dt}\left(\frac{H}{\dot{\bar{\varphi}}}\delta\varphi+\Phi \right),\quad
\frac{d}{dt}\left(\frac{a\Phi}{H}\right)=\frac{a\dot{\bar{\varphi}}^2c_s^2}{2H^2}\left(\frac{H}{\dot{\bar{\varphi}}}\delta\varphi+\Phi \right),
\end{eqnarray}
and they differ from the classical ones in the square of the velocity of sound that appears in the right hand side of the second equation, and
whose value is
\begin{eqnarray}
c_s^2\equiv {\Omega}={1-\frac{2\rho}{\rho_c}}.\end{eqnarray}

Introducing the variables
\begin{eqnarray}
 v=a(\delta\varphi+\frac{\dot{\bar{\varphi}}}{{H}}\Phi); \quad z=\frac{a\dot{\bar{\varphi}}}{{H}},\quad u=\frac{2\Phi}{\dot{\bar{\varphi}}c_s};
\quad\theta=\frac{1}{c_sz},
\end{eqnarray}
one obtains the corresponding M-S equations (\ref{1}) in holonomy corrected LQC.

Note that in the super-inflationary phase, i.e. when $\rho>\rho_c/2$, the velocity of the sound becomes imaginary which could lead, during this stage,
 to a Jeans instability for ultra-violet modes satisfying $k^2 |c_s^2|\gg \left|\frac{z''}{z}\right|$, and as a consequence,
 these growing modes could condensate an produce undesirable cosmological consequences. This is a problem that needs to be addressed, because the validity of the linear
 perturbation equations during this regime is not clear.

This  is one of the reasons why a {\it 
Teleparallel version of LQC} has recently been introduced in \cite{aho13}. This theory  is based in the fact that, in the flat FLRW geometry, the holonomy
corrected Friedmann equation introduced
above could be obtained as a particular case of a teleparallel $F(T)$ theory.
 In \cite{aho13} this example has been found to be
 \begin{eqnarray}\label{teleparallelLQC}
 F_{\pm}(T)=\pm\sqrt{-\frac{T\rho_c}{2}}\arcsin\left(\sqrt{-\frac{2T}{\rho_c}}\right)+  \frac{\rho_c}{2}\left(1\pm\sqrt{1+\frac{2T}{\rho_c}}  \right),
\end{eqnarray}
where $+$ corresponds to the super-inflationary phase, i.e. to $\rho>\rho_c/2$, and $-$ to the deflationary one, i.e. to $\rho<\rho_c/2$.

\begin{remark}
The function (\ref{teleparallelLQC}) is easily obtained, isolating $\rho$ as a funtion of $T$ in equation  (\ref{friedmann}) and inserting it  in the general Friedmann equation for $F(T)$ gravity
$\rho=-2\frac{d F(T)}{dT}+F(T)$ \cite{aho13}. Note that, in this case, in order to obtain (\ref{teleparallelLQC}) it is not necessary that the critical density be given by the expression 
(\ref{criticaldensity}). It can be understood just as a parameter in the theory.
\end{remark}

The perturbation equations were recently obtained in \cite{h13} using the well-known perturbed
equations in $F(T)$ gravity \cite{ccdds11} applied to this particular $F(T)$ model. The result is
\begin{eqnarray}
 \frac{c_s^2}{a^2}\Delta\Phi=\frac{\dot{\bar{\varphi}}^2\Omega}{2H}\frac{d}{dt}\left(\frac{H}{\dot{\bar{\varphi}}}\delta\varphi+\Phi \right),\quad
\frac{d}{dt}\left(\frac{a\Phi}{H}\right)=\frac{a\dot{\bar{\varphi}}^2\Omega}{2H^2}\left(\frac{H}{\dot{\bar{\varphi}}}\delta\varphi+\Phi \right),
\end{eqnarray}
where, in this teleparallel version of LQC,
the square of the velocity of sound is
\begin{eqnarray} c_s^2\equiv |\Omega|
\frac{\arcsin\left(2\sqrt{\frac{{3}}{\rho_c}}H\right)}{2\sqrt{\frac{3}{\rho_c}}H},
\end{eqnarray}
which is always positive.

Performing the change of variables
\begin{eqnarray} \label{tele} v=a\frac{\sqrt{|\Omega|}}{{c}_{s}}(\delta\varphi+\frac{\dot{\bar{\varphi}}}{{H}}\Phi);\quad
z=\frac{a\sqrt{|\Omega|}\dot{\bar{\varphi}}}{{c}_{s}{H}},\quad u=\frac{2\Phi}{\dot{\bar{\varphi}}\sqrt{|\Omega|}};
\quad\theta=\frac{1}{{c}_{s}z},
\end{eqnarray}
one obtains the following M-S equations in teleparallel LQC, that differ a little bit from (\ref{1})

\begin{eqnarray}\label{1a}
 \frac{|\Omega|}{\Omega}{c}_{s}\Delta u= z\left(\frac{v}{z}\right)';\quad \frac{|\Omega|}{\Omega}\theta\left(\frac{u}{\theta}\right)'= {c}_{s}v.
\end{eqnarray}

Note that, since
in this version of LQC the velocity of sound is always positive, modes that satisfy $k^2 |c_s^2|\gg \left|\frac{z''}{z}\right|$ are
sound waves which never condensate, and thus, they  will not produce any cosmological consequence.

\begin{remark}
Here, it is important to realize that holonomy corrected LQC and what we call {\it teleparallel LQC} only coincide in the homogeneous and isotropic case, i.e., in the flat FLRW geometry. When one
deals with cosmological perturbation the theories are completely different, and lead to different perturbed dynamical equations, because these equations are obtained using
different approaches: In holonomy corrected LQC the perturbation equations are obtained working in the Hamiltoninan framework, where the isotropic part of the Ashtekar connection, 
which does not have a quantum version in the Hilbert space of the almost periodic functions, has to be replaced by a suitable sinus. After this replacement, the anomalies that will appear 
in the algebra of constrains must be removed introducing some counter-terms. On the other hand, the perturbed equations in the teleparallel version are obtained in the Lagrangian
framework using the well-know equations in $F(T)$ gravity.
\end{remark}

\section{Calculation of the  Bardeen potential}
In this Section, we will show that the formula that relates the Bardeen potential with the curvature fluctuation in co-moving coordinates, in the matter bounce scenario,
is the same as the one obtained
in inflationary cosmology
when the Universe enters in the matter dominated phase. In fact, we will see that, the formula only depends of the fact  that, at late times, the Universe is in a matter dominated phase.

To prove this, first of all we
perform the Laplacian in the second equation of (\ref{1}) and use the first one, to get the M-S equation
\begin{eqnarray}\label{2}
 v''-{c}^2_{s}\Delta v-\frac{z''}{z}v=0.
\end{eqnarray}

Now,
 inserting the second equation of (\ref{1}) in the first one, one gets
\begin{eqnarray}\label{3}
 u''-{c}^2_{s}\Delta u-\frac{\theta''}{\theta}u=0.
\end{eqnarray}

\begin{remark} The same happens with equations (\ref{1a}), that is, in {\it the teleparallel version of LQC}  the M-S equations (\ref{2}) and  (\ref{3})
also remain valid. Effectively, performing the Laplacian in the first equation and using the second one, one gets
\begin{eqnarray}
 \frac{|\Omega|}{\Omega}\theta\left(\frac{\Omega}{|\Omega|}z^2\left(\frac{v}{z}\right)'\right)'={c}_{s}\Delta v.
\end{eqnarray}

Now, using that
\begin{eqnarray}
 \frac{|\Omega|}{\Omega}\left(\frac{\Omega}{|\Omega|}z^2\left(\frac{v}{z}\right)'\right)'=\left(z^2\left(\frac{v}{z}\right)'\right)'=v''z-z''v,
\end{eqnarray}
one finally obtains
\begin{eqnarray}
 \theta(v''z-z''v)={c}_{s}\Delta v\Longleftrightarrow v''-{c}_{s}^2\Delta v-\frac{z''}{z}v=0.
\end{eqnarray}

In the same way, inserting the second equation of (\ref{1a}) in the first one, one gets
\begin{eqnarray}
 c_s\Delta u=
 z\frac{\Omega}{|\Omega|}\left(\frac{|\Omega|}{\Omega}\theta^2\left(\frac{u}{\theta}\right)'\right)'
 \Longleftrightarrow
 c_s\Delta u=
 z\left(\theta^2\left(\frac{u}{\theta}\right)'\right)',
\end{eqnarray}
which is equivalent to
\begin{eqnarray}
 c_s\Delta u=
 z(u''\theta-\theta'' u) \Longleftrightarrow u''-{c}_{s}^2\Delta u-\frac{\theta''}{\theta}u=0.
 \end{eqnarray}
\end{remark}

Coming back to equation (\ref{2}),
we will obtain an equivalent integral equation. In Fourier space, we write (\ref{2}) as follows:
\begin{eqnarray}\label{4}
 v_k''-\frac{z''}{z}v_k=-k^2c_s^2v_k,
\end{eqnarray}
and find the solution for the ``associate  equation'', namely $v_k''-\frac{z''}{z}v_k=0$ . Finally, we  use the method of variation of constants to find a particular solution.
The solution of the ``associate  equation''  is
\begin{eqnarray}\label{5}
 v_{k}(\eta)=A_1(k)z(\eta)+A_2(k)z(\eta)\int^{\eta}\frac{d\bar{\eta}}{z^2(\eta)},
\end{eqnarray}
and  the method of variation of constants, after some algebra,  gives the following solution of equation (\ref{4}) as an integral equation
\begin{eqnarray}\label{6}
 v_{k}(\eta)=A_1(k)z(\eta)+A_2(k)z(\eta)\int^{\eta}\frac{d\bar{\eta}}{z^2(\eta)} \nonumber\\
-k^2z(\eta)\int^{\eta}\frac{d\bar{\eta}}{z^2(\bar{\eta})}\int^{\bar{\eta}}z(\tilde{\eta})c_s^2(\tilde{\eta})v_k(\tilde{\eta})d\tilde{\eta}.
\end{eqnarray}

To obtain the corresponding integral equation for $u_k$ we use the first equation of (\ref{1}), $-c_sk^2u_k=z\left(\frac{v_k}{z}\right)'$. Inserting in it
the expression (\ref{6}) one gets
\begin{eqnarray}\label{7}
 u_{k}(\eta)=-\frac{A_2(k)}{k^2}\theta(\eta)+
\theta(\eta)\int^{\eta}\frac{c_s(\bar{\eta})}{\theta(\bar{\eta})}v_k(\bar{\eta})d\bar{\eta}.
\end{eqnarray}

From this formula we can calculate the power spectrum for the Bardeen potential
at late times when the Universe is matter dominated. To do that, we consider modes well outside of the Hubble radius, i.e., modes
that satisfy $c_s^2k^2\ll 1/\eta^2\sim a^2H^2\sim z''/z$. In that case  formula (\ref{5}) becomes
\begin{eqnarray}\label{b1}v_k(\eta)\cong A_1(k)z(\eta),\end{eqnarray}
that is, the curvature fluctuation in co-moving coordinates, defined as
\begin{eqnarray}
\zeta_k(\eta)\equiv\frac{v_k(\eta)}{z(\eta)},\end{eqnarray} is constant.

Then, since at late times $\theta(\eta)\propto \frac{1}{\eta^2}$, this means that $\theta(\eta)$ is the decaying mode, and thus, inserting (\ref{b1}) into
(\ref{7})  one obtains
\begin{eqnarray}\label{9}
 u_{k}(\eta)={\zeta_k}
\theta(\eta)\int^{\eta}\frac{d\bar{\eta}}{\theta^2(\bar{\eta})}.
\end{eqnarray}

Now,
taking into account that,  for a matter dominated Universe, in  EC one has
$z(\eta)= \sqrt{3}a(\eta)$,
and using the following classical relations
\begin{eqnarray}\label{11}
 \Phi_k=\frac{u_k}{2}\sqrt{\rho};\quad \rho(t)=\frac{4}{3t^2};\quad t\propto\eta^3,
\end{eqnarray}
one finally obtains the relation between the Bardeen potential and the curvature fluctuation in co-moving coordinates
\begin{eqnarray}\label{12}
 \Phi_k(\eta)=\frac{3}{5}\zeta_k(\eta).
\end{eqnarray}

Moreover, in EC one of the perturbation equations is
\begin{eqnarray}\label{c1}
 -\frac{k^2}{a^2}\Phi_k-3H\dot{\Phi}_k-3H^2\Phi_k=\frac{1}{2}\delta\rho_k,
\end{eqnarray}
and since for modes well outside the Hubble radius ($k^2\ll a^2H^2$) $\Phi_k$ is constant, the density contrast
$\delta_k\equiv \frac{\delta\rho_k}{\rho}$  is related with the
curvature in co-moving coordinates  as follows
\begin{equation}\label{c2}
 \delta_k=-2\Phi_k=-\frac{6}{5}\zeta_k(\eta).
\end{equation}

Note the remarkable fact that  relations  (\ref{12})
only depend of the fact that, at late times, the Universe obeys EC and is matter dominated.
That happens in inflation
when one one considers a transition form the de Sitter phase to the matter domination (see for instance \cite{r02})
and in the matter bounce scenario. Then, our result complements the
duality  pointed out in \cite{w99} where was showed that in EC
the de Sitter phase, where $a(\eta)\propto -\frac{1}{\eta}$, and the matter-domination $a(\eta)\propto \eta^2$
lead to the same equation $(\ref{2})$
\begin{eqnarray}\label{a}
 v''-\Delta v -\frac{2}{\eta^2}v=0,
\end{eqnarray}
meaning that de Sitter inflation and the matter bounce scenario give rise to a scale invariant spectrum for
the curvature fluctuation in co-moving coordinates.
In fact,  in next section we will calculate  the value  of the arbitrary function $A_1(k)$ that appears in (\ref{b1}), and we will show that  is of the order $k^{-3/2}$ 
(see formulas (\ref{21}) and (\ref{23})).

Now, from the relation (\ref{12}) and the fact that in the matter bounce bounce scenario the power spectrum of the curvature fluctuation
in co-moving coordinates is scale invariant (see next Section)
one can conclude that, in the matter
bounce scenario, the Bardeen potential is also scale invariant.

\section{Calculation of the power spectrum of scalar perturbations in LQC}
In this section we perform an  study of the way to calculate analytically the power spectrum of the curvature fluctuation in co-moving coordinates,
and from (\ref{c2}) the density contrast,
when one considers the matter bounce scenario in holonomy corrected LQC and in its teleparallel version, which provides the easiest model to calculate analytically it.

Solving the holonomy corrected Friedmann equation in the flat FLRW spacetime and the conservation equation for a matter dominated Universe
\begin{eqnarray}
 H^2=\frac{\rho}{3}\left(1-\frac{\rho}{\rho_c}\right);\quad \dot{\rho}=-3H\rho,
\end{eqnarray}
one obtains the following quantities \cite{h12}
\begin{eqnarray}\label{13}
a(t)=\left(\frac{3}{4}\rho_ct^2+1\right)^{1/3}, \quad H(t)=\frac{\frac{1}{2}\rho_ct}{\frac{3}{4}\rho_ct^2+1}\quad\mbox{and}\quad
\rho(t)=\frac{\rho_c}{\frac{3}{4}\rho_ct^2+1}.
\end{eqnarray}

For small values of the energy density ($\rho\ll \rho_c$), EC is recovered and equation (\ref{2}) becomes the usual M-S
equation that for a matter-dominated Universe, working in Fourier space,  is given by
\begin{eqnarray}\label{14}
 v_k''+\left(k^2 -\frac{a''}{a}\right)v_k=0\Longleftrightarrow
 v_k''+\left(k^2 -\frac{2}{\eta^2}\right)v_k=0.
\end{eqnarray}

Assuming that at early times the Universe is in the Bunch-Davies (adiabatic) vacuum, one must take for $\eta\rightarrow-\infty$
the following mode function
\begin{eqnarray}\label{15}
 v_k(\eta)=\frac{e^{-ik\eta}}{\sqrt{2k}}\left(1-\frac{i}{k\eta}\right).
\end{eqnarray}

At early times all the modes are inside the Hubble radius, and when time moves forward the modes leave this radius. Then, for a matter-dominated Universe
in EC, the modes well outside the Hubble radius are characterized by the long wavelength condition
\begin{eqnarray}\label{16}
 k^2\eta^2 \ll 1\Longleftrightarrow k^2 \ll \left|\frac{a''}{a}\right|\Longleftrightarrow k^2 \ll \left|\frac{1}{{c}^2_{s}}\frac{z''}{z}\right|,
\end{eqnarray}
because for small values of $\rho$ one recovers EC  where $z=\sqrt{3}a$ and $c_s=1$.

And, when holonomy effects are not important,  for modes well outside the Hubble radius the M-S equation becomes
\begin{eqnarray}\label{17}
 v_k'' -\frac{z''}{z}v_k=0,
\end{eqnarray}
which can be solved using the method of reduction of the order, giving as a result the following long wavelength approximation
\begin{eqnarray}\label{18}
 v_k(\eta)=B_1(k)z(\eta)+B_2(k)z(\eta)\int_{-\infty}^{\eta}\frac{d\bar{\eta}}{z^2(\bar{\eta})},
\end{eqnarray}
where for convenience we have taken a definite integral. The reason why we have made this choice is that, in teleparallel LQC, it is impossible to calculate explicitly
the primitive of $1/z^2(\bar{\eta})$. However, for $\eta\rightarrow -\infty$, if we  make the approximation  $z\cong\sqrt{3}a$ we will obtain
$\int_{-\infty}^{\eta}\frac{d\bar{\eta}}{z^2(\bar{\eta})}\cong\int_{-\infty}^{\eta}\frac{d\bar{\eta}}{3a^2(\bar{\eta})}$, and this last integral can be analytically calculated.

Note that,
 at early times in the contracting phase, for modes well outside the Hubble radius, the expressions (\ref{15}) and (\ref{18}) give the same
solution. The solution given by (\ref{15}) could be expanded in terms of $k\eta\ll 1$, and retaining  the leading terms in the real and
imaginary parts of $v_k$,
one gets
\begin{eqnarray}\label{19}
v_k(\eta)\cong -\frac{k^{3/2}\eta^2}{3\sqrt{2}}-\frac{i}{\sqrt{2}k^{3/2}\eta}.
\end{eqnarray}

On the other hand,
the explicit solution of
(\ref{18}), as we have already explained, is obtained using the approximation  $z\cong\sqrt{3}a=\frac{1}{4\sqrt{3}}\rho_c\eta^2$,
which  gives as a result
\begin{eqnarray}\label{20}
v_k(\eta)\cong\frac{B_1(k)}{4\sqrt{3}}\rho_c\eta^2-\frac{4B_2(k)}{\sqrt{3}\rho_c}\frac{1}{\eta}.
\end{eqnarray}

 Matching both solutions one obtains
 \begin{eqnarray}\label{21}
 B_1(k)= -\sqrt{\frac{8}{3}}\frac{k^{3/2}}{\rho_c}  \quad \mbox{and}\quad B_2(k)=i\sqrt{\frac{3}{8}}\frac{\rho_c}{2k^{3/2}}.
\end{eqnarray}

Once we have calculated the coefficients $B_1(k)$ and $B_2(k)$ we use equation (\ref{18}) to calculate $v_k$ at late times. More precisely,
we calculate   $v_k$  in the classical regime of the expanding phase for modes that are still  well outside of the Hubble radius. Note that
we are considering modes that
in the contracting phase leave the Hubble radius and then evolve
satisfying $ k^2 \ll \left|\frac{1}{{c}^2_{s}}\frac{z''}{z}\right|$. Then, we can use the long wavelength approximation
\begin{eqnarray}\label{22}
 v_k(\eta)=(B_1(k)+B_2(k)R)z(\eta),
\end{eqnarray}
where $R\cong\int_{-\infty}^{\infty}\frac{d\bar{\eta}}{z^2(\bar{\eta})}$, because $\eta$ is large enough.

From (\ref{22}) one has
\begin{eqnarray}\label{23}
 \zeta_k(\eta)=\frac{ v_k(\eta)}{z(\eta)}= B_1(k)+B_2(k)R\cong B_2(k)R ,
\end{eqnarray}
and thus, the scalar power spectrum is given by
\begin{eqnarray}\label{24}
 {\mathcal P}_{\zeta}(k) \equiv
 \frac{k^3}{2\pi^2}|\zeta_k(\eta)|^2
 =\frac{ 3\rho_c^2}{64\pi^2}R^2=\frac{ 3\rho_c^2}{\rho_{pl}}R^2,
\end{eqnarray}
because since in our units $8\pi G=1$, one has $\rho_{pl}=64\pi^2$.

In the case of
 holonomy corrected LQC one has  $z(t)= \frac{2a^{5/2}(t)}{\sqrt{\rho_c}t}$ (see \cite{w13}), which leads to a simple calculation of
$R^2$ giving as a result $\frac{\pi^2}{27\rho_c}$. Consequently, in holonomy corrected LQC one has
\begin{eqnarray}\label{25}
 {\mathcal P}_{\zeta}(k)=\frac{\rho_c}{576}=\frac{\pi^2}{9}\frac{\rho_c}{\rho_{pl}}.
\end{eqnarray}

On the other hand, in teleparallel LQC one has
\begin{eqnarray}\label{26}
 z(t)=2\left(\frac{3}{\rho_c}\right)^{1/4}\frac{a(t)|t|^{1/2}}{t\sqrt{\arcsin\left(\frac{\sqrt{3\rho_c}|t|}{a^3(t)} \right)}},
\end{eqnarray}
giving as a power spectrum
\begin{eqnarray}\label{27}
 {\mathcal P}_{\zeta}(k)
 =\frac{\rho_c}{144\pi^2}\left(
\int_0^{\pi/2}\frac{x}{\sin x}dx \right)^2
= \frac{\rho_c}{36\pi^2}{\mathcal C }^2= \frac{16}{9}\frac{\rho_c}{\rho_{pl}}{\mathcal C }^2,
\end{eqnarray}
where ${\mathcal C}=1-\frac{1}{3^2}+\frac{1}{5^2}-\frac{1}{7^2}+...= 0.915965...$ is Catalan's constant.

The key point to obtain the scale invariant power spectrum  (\ref{25}) (resp. (\ref{27})) is that one only considers
modes that after leaving and before re-entering the Hubble radius, when holonomy effects could be disregarded,
 satisfy the long wavelength condition $k^2 |c_s^2|\ll \left|\frac{z''}{z}\right|$, that is,
the term ${c}^2_{s}\Delta v$  in the M-S
equation is disregarded between the exit and the reentry of the modes in the  Hubble radius.

In holonomy corrected LQC and in its teleparallel version,  since the symmetric function $z''/z$  is increasing for $-\infty<t<0$ and decreasing for $0<t<\infty$, and $|c_s^2|$ satisfy
$|c_s^2|\leq 1$ and $\lim_{t\rightarrow \pm\infty}|c_s^2|=1$, all the modes that leave the Hubble radius at a early time $-|T|$ satisfy
the relation
$k^2 |c_s^2|\ll \left|\frac{z''}{z}\right|$ up to late time $|T|$. Then, we can conclude that formulae (\ref{25}) and (\ref{27}) are correct for all modes that
leave the Hubble radius at early times (when the Universe is in the classical regime).

\begin{remark}
 Dealing with holonomy corrected LQC, where the square of the velocity of sound is $c_s^2=\Omega=1-2\frac{\rho}{\rho_c}$, one has to be cautious because  in the super-inflationary phase
 the speed of sound becomes negative. In this analysis we are only taking into account modes satisfying the long wavelength conditions, and thus, for those modes the speed of sound
 does not have any importance in their evolution. However, and this is a problem that has not been addressed yet, the modes with a shorter wavelength (modes remaining inside the Hubble radius) 
 will suffer Jeans instability, leading
 to some undesirable cosmological consequences. On the other hand, this never happens in the teleparallel version of LQC where the speed of sound is always a real number.
\end{remark}

In contrast, the quantity $\left|\frac{\theta''}{\theta}\right|$  vanishes at the bouncing time. Then, in bouncing scenarios
it is impossible to calculate $u_k$ (and the Bardeen potential $\Phi_k$) using the long wavelength approximation
\begin{eqnarray}\label{28}
 u_k(\eta)=C_1(k)\theta(\eta)+C_2(k)\theta(\eta)\int^{\eta}\frac{d\bar{\eta}}{\theta^2(\bar{\eta})},
\end{eqnarray}
because the relation $k^2 |c_s^2|\ll \left|\frac{\theta''}{\theta}\right|$ after leaving and before re-entering the Hubble radius, doesn't hold
for any mode. Instead of (\ref{28}), in order to calculate $u_k$, one has to use the exact expression given by formula (\ref{7}).

However, in inflationary EC formula (\ref{28}) leads to the correct power spectrum for the Bardeen potential.
Effectively, the vacuum state
is given by the modes (\ref{15}), which together with the first equation of (\ref{1}) allow us to calculate the modes $u_k$ at early times. Then, for modes
well outside  the Hubble radius ($|k\eta|\ll 1$) one has
\begin{eqnarray}
u_k(\eta)\cong \frac{1}{\sqrt{2k}}\eta-\frac{i}{\sqrt{2}k^{3/2}}.
\end{eqnarray}

This expression has to be matched with (\ref{28}) during the quasi de Sitter phase. Since in  EC one has $z=\frac{a\sqrt{-2\dot{H}}}{H}$, during
the quasi de Sitter epoch we can approximate $z$ by $-\frac{\sqrt{\epsilon}}{H\eta}$ where $H$ and $\epsilon\equiv -\frac{2\dot{H}}{H^2}$ could be
considered constants. Then, a simple calculation gives rise to
\begin{eqnarray}
C_1(k)=-\sqrt{\frac{\epsilon}{2k}}\frac{1}{H},\quad C_2(k)=\frac{-iH}{\sqrt{2\epsilon}k^{3/2}}.
\end{eqnarray}

Now, to calculate the Bardeen potential at late times, we use the classical relations
$\Phi_k=\frac{\sqrt{-2\dot{H}}}{2a}u_k$ and $\theta=\frac{1}{z}$ to obtain
\begin{eqnarray}
\Phi_k(t)=\frac{C_1(k)H(t)}{2a(t)}-\frac{C_2(k)H(t)}{a(t)}\int^t\frac{a(\bar{t})\dot{H}(\bar{t})}{H^2(\bar{t})}d\bar{t}.
\end{eqnarray}

The first term is decaying and can be disregarded. The second one, after integration by parts, leads to

\begin{eqnarray}
\Phi_k(t)= C_2(k)\frac{d}{dt}\left(\frac{1}{a(t)}\int^ta(\bar{t})d\bar{t}\right).
\end{eqnarray}

When the Universe is matter dominated, i.e., when $a(t)\propto t^{2/3}$ one obtains
\begin{eqnarray}
\Phi_k(t)= \frac{3}{5}C_2(k).
\end{eqnarray}

Finally,  calculating  (\ref{5}) in the quasi de Sitter phase and matching the result with (\ref{19})  one easily obtains $A_1(k)=C_2(k)$, and taking
into account that the mode
$z(t)\int^t\frac{d\bar{t}}{z^2(\bar{t})}$ is decaying in the matter dominated stage one concludes that $\zeta_k(t)=C_2(k)$, and thus,
we obtain in inflationary cosmology the relation (\ref{12}).

\section{The current model}
To calculate the power spectrum  provided by LQC  in the matter bounce scenario, first of all one has to look for a potential of the scalar field such that
its non-perturbed solutions (the background solutions) lead to a matter dominated Universe, i.e., they depict, at very early times, a
matter dominated Universe in the contracting phase that evolves towards a bounce to enter in the expanding phase.

The simplest way
to find one such potential is to impose that the pressure vanishes, i.e., $\frac{\dot{\bar\varphi}^2}{2}-V(\bar\varphi)=0$, which
leads to the equation
\begin{eqnarray}
 \dot{\bar\varphi}^2(t)=\rho(t)\Longleftrightarrow \dot{\bar\varphi}^2(t)=\frac{\rho_c}{\frac{3}{4}\rho_ct^2+1},
\end{eqnarray}
where we have used the third equation of (\ref{13}).

This equation has the particular solution
\begin{eqnarray}\label{sol}
 \bar\varphi(t)=\frac{2}{\sqrt{3}}\ln\left(\sqrt{\frac{3}{4}\rho_c} t+\sqrt{\frac{3}{4}\rho_c t^2+1}  \right),
\end{eqnarray}
which, after reconstruction, i.e., isolating $\frac{3}{4}\rho_ct^2+1$ as a function of $\bar\varphi$ and using the relation $\f{\rho_c}{\frac{3}{4}\rho_ct^2+1}=2V(\bar\varphi)$, leads to the potential
\begin{eqnarray}\label{pot}
 V({\varphi})=2\rho_c\frac{e^{-\sqrt{3}{\varphi}}}{\left(1+e^{-\sqrt{3}{\varphi}}\right)^2}.
\end{eqnarray}

It is important to realize that the
 solution (\ref{sol}) is special in the sense that it satisfies for all time $\dot{\bar\varphi}^2(t)/2=V(\bar\varphi(t))$, that is,
 if the Universe is described by this solution it will be  matter dominated dominated all the time. However, the other
solutions, that is, the solutions of the non-perturbed conservation equation
\begin{eqnarray}\label{KG}
 \ddot{\bar{\varphi}}+3{H}_\pm\dot{\bar{\varphi}}+{ {V}_{\varphi}(\bar{\varphi})}=0,
\end{eqnarray}
where ${H}_- =-\sqrt{\frac{{\rho}}{3}(1-\frac{\rho}{\rho_c})}$ in the contracting phase  and
${H}_+ =\sqrt{\frac{{\rho}}{3}(1-\frac{\rho}{\rho_c})}$ in the expanding phase,
do not lead to a matter-dominated Universe all the time. Only at early and late times the Universe is matter dominated because the solution (\ref{sol}) is a global repeller
at early times and
a global attractor at late times. But, what is important to realize is that all these solutions depict a matter bounce scenario: matter domination at early times in the contracting phase,
evolution towards a bounce and finally entrance in the expanding phase.

The method to prove the asymptotical behavior of all the solutions is similar to the one used in \cite{hae}, and goes as follows:

{ At early and late times one can disregard holonomy corrections, and 
since we are considering the early and late time dynamics of the system (what happens for large values of $|\varphi|$),  our potential (\ref{pot}) reduces to 
$\bar{V}(\varphi)=V_0e^{-\sqrt{3}\varphi}$.
Then, performing the change of variable $\bar{\varphi}=\frac{2}{\sqrt{3}}\ln \psi$ the corresponding  non-perturbed Klein-Gordon equation (\ref{KG})
(or equivalently,
the conservation equation) reads
\begin{eqnarray}\label{eq1}
 \frac{d\dot{\psi}}{d{\bar\varphi}}=F_{\pm}(\dot{\psi}),
\end{eqnarray}
with
\begin{eqnarray}
 F_{\pm}(\dot{\psi})=\frac{3\sqrt{3}}{4\dot{\psi}}\left(\frac{2}{3}\dot{\psi}^2+V_0 \right)\mp\frac{3}{2}\sqrt{\frac{2}{3}\dot{\psi}^2+V_0},
\end{eqnarray}
where in $F_{\pm}$, the sign $+$ (resp. $-$) means that the Universe is in the expanding (resp. contracting) phase.

The equation (\ref{eq1}) describes two (one for the contracting and other one for the expanding phase) one dimensional first order autonomous dynamical systems, that are completely
understood
calculating its critical points
and evaluating the function $F_{\pm}$ at the right and left hand sides of each critical point in order to know whether these critical points are attractors or repellers.
In our case, these critical points are $\dot{\psi}_{+}=\sqrt{\frac{3}{2}V_0}$ for the expanding phase and  $\dot{\psi}_{-}=-\sqrt{\frac{3}{2}V_0}$
for the contracting one. Taking into account the sign of $F_{\pm}$ in the right and left hand sides of each critical point one will deduce the dynamics of the system
in time $\bar\varphi$. For example, in figure $1$ we show the phase portrait
(in the straight line $\dot{\psi}$)
when the Universe is in the contracting phase,
and from the phase portrait in time $\bar\varphi$, we will deduce the corresponding one in cosmic time, changing the direction of the arrows for negative values of $\dot{\psi}$
(see figure $2$), because  $\dot{\psi}<0$ which is the same that $\dot{\bar\varphi}<0$, means that $\bar\varphi$ is decreasing in cosmic time, then in order to obtain the dynamics in
cosmic time (figure $2$), the direction of the arrows must
be changed in the phase portrait $1$.

\begin{figure}
\begin{eqnarray*}
\begin{picture}(74,44)\thicklines
\put(-40,-10){\line(1,0){240}}
\put(-80,-10){\line(1,0){100}}
\put(30,-10){\vector(-1,0){2}}
\put(-60,-10){\vector(1,0){2}}
\put(120,-10){\vector(1,0){2}}
\put(-20,-10){\circle*{3}}
\put(50,-10){\circle*{3}}
\put(-20,-14){\makebox(0,0)[t]{$-\sqrt{\frac{3}{2}V_0}$}}
\put(50,-14){\makebox(0,0)[t]{$0$}}
\end{picture}
\end{eqnarray*}
\vspace{0.25cm}

\caption{Phase portrait, in the contracting phase, using $\bar\varphi$ as a time. With this time, $0$ is a repeller and $-\sqrt{\frac{3}{2}V_0}$ an attractor.}
\end{figure}
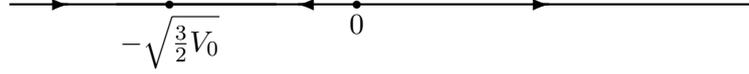

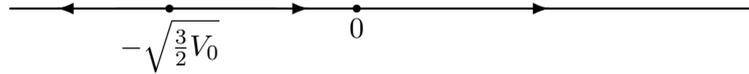
\begin{figure}
\begin{eqnarray*}
\begin{picture}(74,44)\thicklines
\put(-40,-10){\line(1,0){240}}
\put(-80,-10){\line(1,0){100}}
\put(30,-10){\vector(1,0){2}}
\put(-60,-10){\vector(-1,0){2}}
\put(120,-10){\vector(1,0){2}}
\put(-20,-10){\circle*{3}}
\put(50,-10){\circle*{3}}
\put(-20,-14){\makebox(0,0)[t]{$-\sqrt{\frac{3}{2}V_0}$}}
\put(50,-14){\makebox(0,0)[t]{$0$}}
\end{picture}
\end{eqnarray*}
\vspace{0.25cm}

\caption{ Phase portrait, in the contracting phase, using the cosmological time $t$. Now
 $-\sqrt{\frac{3}{2}V_0}$ is a global repeller.}
\end{figure}

The conclusion is that in the expanding (resp. contracting) phase $\dot{\psi}_{+}$ (resp. $\dot{\psi}_{-}$) is  a global attractor (resp. repeller). These points correspond to
the solutions $\bar\varphi_{\pm}=\frac{2}{\sqrt{3}}\ln(\pm \sqrt{\frac{3}{2}V_0}t)$, that of course, satisfy $\rho=\dot{\bar\varphi}^2=\frac{4}{3t^2}$, i.e., all the solutions of
the conservation equation depict a  
matter-dominated Universe
at early  (in the contracting phase) and  late times (in the expanding one).}

\begin{figure}[h]
\begin{center}
\includegraphics[scale=0.35]{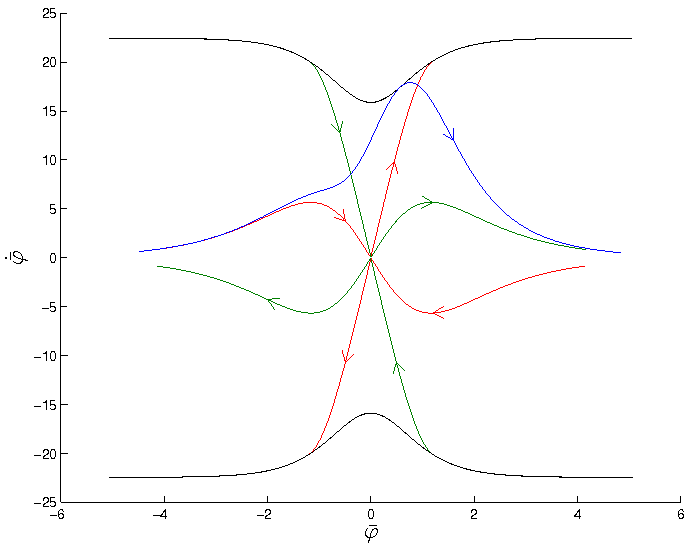}
\includegraphics[scale=0.35]{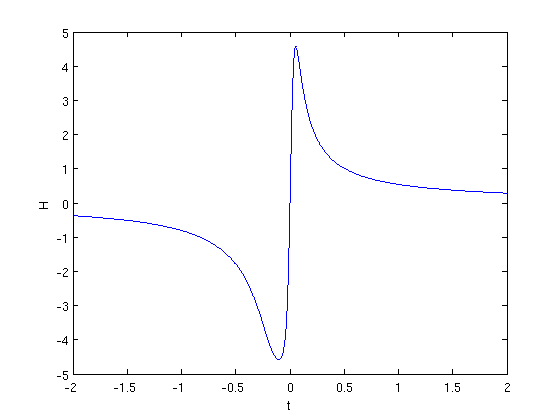}
\end{center}

\caption{{\protect\small In the first picture we have the phase portrait:  black curves   defined by $\rho=\rho_c$ depict the points where the Universe bounces. The point $(0,0)$ is a
saddle point, red (resp. green) curves are the invariant
curves  in the contracting (resp. expanding) phase. The blue curve corresponds to an orbit different from the analytically computed one (\ref{sol}).
Note that, before (resp. after) the bounce the blue curve does not cut the red (resp. green) curves. Finally, it is important to realize that the allowed orbits are those that touch the
black curve in the region delimited by an unstable red curve and a stable green curve, because for  orbits that do not satisfy this condition, $\dot{\bar\varphi}$ vanishes at some time,
meaning that its corresponding power spectrum diverges.
In the second picture we have drawn the Hubble parameter for the blue curve of the first picture.}}
\end{figure}

Then, once we have proved the behavior of the solutions at early and late times,  to calculate analytically the power spectrum
we need the evolution of $a(t)$, $\bar\varphi(t)$ and $H(t)$ during all the time, and we only have, analytically, this evolution for the particular solution (\ref{sol}).  For
the other solutions, numerical calculations are needed
(In figure 3 we show the phase portrait in the plane $(\bar\varphi,\dot{\bar\varphi})$ for the dynamical system given by equation (\ref{KG})).
In fact, using this analytical solution, to obtain a theoretical value of the power spectrum that
matches correctly with observations, one can see that the value of the critical energy density has to be of the order
$10^{-9}\rho_{pl}$ (see \cite{w13}). This does not mean, contrary to the claim of \cite{h13,w13}, that the correct value of the critical density has to be of this order.  What it
really means is  that the value of the critical density has to be of this order if
the Universe is described by  orbits
near this analytical one, but there could be other orbits such that for the current value of the critical density, approximately  $0.4 \rho_{pl}$, the numerical results obtained from
the model might match
correctly with current observations
(we will show at the end of the Section that this is not the case, but it could be possible in principle). One has to imagine $\rho_c$ as a parameter, whose value depends on the
orbit we have chosen, and  is determined by current observations. This is
exactly what happens in chaotic inflation. More precisely, if one considers, for the sake of simplicity, the quadratic model $V(\varphi)=\frac{1}{2}m^2\varphi^2$ the number of e-folds
before the end of inflation is
approximately
\begin{eqnarray}
 N=\int_{t_N}^{t_{end}}H(t)dt\cong \int_{\bar\varphi_{end}}^{\bar\varphi_{N}}\frac{V(\bar\varphi)}{{ V_{\varphi}(\bar\varphi)}}d\bar\varphi\cong
 \frac{\bar\varphi^2_N}{4}.
\end{eqnarray}


On the other hand, the power spectrum of scalar perturbations, in slow-roll inflation, is given by
\begin{eqnarray}\label{inflation}
 {\mathcal P}_{\zeta}(k)=\frac{V^3(\bar{\varphi}_N)}{12\pi^2 V_{\varphi}^2(\bar{\varphi}_N)}=
 \frac{m^2\bar{\varphi}_N^4}{96\pi^2}=\frac{32N^2m^2}{3\rho_{pl}}.
\end{eqnarray}

Then, to choose the number of e-folds is equivalent to choosing the orbit, because with the value of $N$ one obtains $\bar\varphi_N$, and with the slow-roll equation
$V(\bar\varphi_N)\dot{\bar\varphi}_N+ V_{\varphi}(\bar\varphi_N)=0$ one obtains $\dot{\bar\varphi}_N$. Finally, for a given value of $N$, using the current constrain
${\mathcal P}_{\zeta}(k)\simeq 2\times 10^{-9}$ and the formula (\ref{inflation}) one determines de value of the parameter $m$, to match correctly theoretical results with
observational data.

Coming back to
 LQC, in the matter bounce scenario the power spectrum is given by equation  (\ref{24})
\begin{eqnarray}\label{powerspectrum}
 {\mathcal P}_{\zeta}(k)
 =\frac{ 3\rho_c^2}{\rho_{pl}}R^2,
\end{eqnarray}
with $R\cong\int_{-\infty}^{\infty}\frac{d{\eta}}{z^2({\eta})}=\int_{-\infty}^{\infty}\frac{dt}{a(t)z^2(t)}$, where in holonomy corrected LQC  $z=\frac{a\dot{\bar{\varphi}}}{H}$, and in
teleparallel LQC $z$ is given by
formula (\ref{tele}).
It is important to realize that, when we have done the matching between equations (\ref{19}) and (\ref{20}) we have used, at early times, the following
scale factor $a(t)\cong \left(\frac{3}{4}\rho_ct^2\right)^{1/3}$. Then, to perform numerical calculations with formula (\ref{powerspectrum}),  one has to use, as a
scale factor,
the solution of $\frac{\dot{a}}{a}=H$ that, at early times, satisfies $a(t)\cong \left(\frac{3}{4}\rho_ct^2\right)^{1/3}$.

Finally,
performing the change of variable $\tilde{t}=\sqrt{\rho_c}t$, we can see that the conservation equation for the homogeneous part of the field (\ref{KG}) becomes
\begin{eqnarray}
 \frac{d^2\bar{\varphi}}{d\tilde{t}^2}+3\tilde{H}_\pm\frac{d\bar{\varphi}}{d\tilde{t}}+{ \tilde{V}_{\varphi}(\bar{\varphi})}=0,
\end{eqnarray}
with $\tilde{V}=\frac{V}{\rho_c}$ and $\tilde{H}_\pm =\frac{H_{\pm}}{\sqrt{\rho_c}}=\pm\sqrt{\frac{\tilde{\rho}}{3}(1-\tilde{\rho})}$, being
$\tilde{\rho}=\frac{\rho}{\rho_c}=\frac{1}{2}\left( \frac{d\bar{\varphi}}{d\tilde{t}}\right)^2+\tilde{V}$.

This means that $\frac{\dot{\bar{\varphi}}(t)}{H(t)}=\frac{1}{\tilde{H}(\tilde{t})}\frac{d\bar{\varphi}(\tilde{t}) }{d\tilde{t}}$ does not depend on $\rho_c$. In the same way,
$a(\tilde{t})$ does not depend on $\rho_c$ because it satisfies the equation $\frac{1}{a(\tilde{t})}\frac{da(\tilde{t})}{d\tilde{t}}=\tilde{H}(\tilde{t})$, and from the definition of the
velocity of sound we see that $c_s(\tilde{t})$ is  independent on the critical density. From this, we can conclude that $z(\tilde{t})$ does not depend on the value of the
critical density, and then
\begin{eqnarray}
 {\mathcal P}_{\zeta}(k)
 =\frac{ 3\rho_c}{\rho_{pl}}\left( \int_{-\infty}^{\infty}\frac{d{\tilde{t}}}{a(\tilde{t})z^2(\tilde{t})} \right)^2,
\end{eqnarray}
which is of the order $\rho_c$. Finally, depending on the chosen orbit, one will numerically obtain different values of $\left( \int_{-\infty}^{\infty}\frac{d{\tilde{t}}}{a(\tilde{t})z^2(\tilde{t})} \right)^2$,
 which determine the corresponding value of the critical density
using the constraint ${\mathcal P}_{\zeta}(k) \cong 2\times 10^{-9}$.

\subsection{Some comments about  the current model}
An important  difficulty of the model given by
 the potential (\ref{pot}) is that it
 cannot explain the current acceleration of the Universe. Moreover, in previous studies
 any mechanism to  explain the reheating of the Universe creating light particles
 like  an oscillatory behavior of the  field  \cite{kls}, an instant preheating \cite{fkl}
or a phase transition to a quasi de Sitter stage
to a radiation dominated Universe \cite{peebles}, has not taken into account this acceleration.

\vspace{0.5cm}

Another problem appears  when one deals with tensor perturbations.
In the case of holonomy corrected LQC this equation is
\begin{eqnarray}
 v_T''-c^2_s\Delta v_T-\frac{z_T''}{z_T}v_T=0,
\end{eqnarray}
where $c_s^2=\Omega$ and $z_T\equiv {a }{{\Omega}}^{-1/2}$. Note that $z_T$ becomes imaginary in the super-inflationary phase $\rho\in(\rho_c/2,\rho_c]$. Moreover, this equation is
singular when $\rho=\rho_c/2$, then there are infinitely many ways to match solutions at this value, and consequently
infinitely many modes could be used to calculate the power spectrum of tensor perturbations giving completely different results (see \cite{h13}). On the other hand, in teleparallel
LQC the corresponding M-S equation for tensor perturbations does not contain singularities but, when one uses the analytic solution (\ref{sol}), the ratio of tensor to scalar perturbations
which is given by
\begin{eqnarray}\label{ratio}
r\cong \frac{1}{3}\left(\frac{\int_{-\infty}^{\infty}\frac{1}{z_T^{2}}d\eta}{\int_{-\infty}^{\infty}\frac{1}{z^{2}}d\eta}\right)^2=
\frac{1}{3}\left(\frac{\int_{-\infty}^{\infty}\frac{1}{az_T^{2}}dt}{\int_{-\infty}^{\infty}\frac{1}{az^{2}}dt}\right)^2,
\end{eqnarray}
where now $z_T\equiv \frac{a{c}_s}{\sqrt{|\Omega|}}$,
is
$r=3\left(\frac{Si(\pi/2)}{{\mathcal C}}\right)^2\cong 6.7187$, as $Si(x)\equiv\int_0^x\frac{\sin y}{y}dy$
is the Sine integral function \cite{ha14}, which is in
contradiction with the current observational bound. Recall that BICEP2 data constrain  this ratio to $r=0.20^{+0.07}_{-0.05}$ with $r=0$ disfavored at $7.0 \sigma$ \cite{Kovac},
and {\it Planck's} data bounds this ratio to be $r<0.11$ (95 \% CL) \cite{Ade}.

\begin{remark}
 Note that our correct definition of $z_T$ differs from the one of \cite{h13} by the factor $\sqrt{2}$. In fact, for small values of the energy density one has $z_T\cong a$ which
 coincides with the classical definition of $z_T$, and which does not happen if one uses the definition given in \cite{h13}. On the other hand, to obtain the formula
 (\ref{ratio}), one has to follow the same steps as in Section 4.  Obtaining the formula
 \begin{eqnarray}
   v_{T,k}(\eta)=B_{T,1}(k)z_T(\eta)+B_{T,2}(k)z_T(\eta)\int_{-\infty}^{\eta}\frac{d\bar{\eta}}{z_T^2(\bar{\eta})},
 \end{eqnarray}
 with
 \begin{eqnarray}
 B_{T,1}(k)= -\sqrt{{8}}\frac{k^{3/2}}{\rho_c}  \quad \mbox{and}\quad B_{T,2}(k)=i\sqrt{\frac{1}{8}}\frac{\rho_c}{2k^{3/2}}.
\end{eqnarray}

Then, the power spectrum for tensor perturbations is
\begin{eqnarray}\label{powertensor}
 {\mathcal P}_{T}(k) \equiv
 \frac{k^3}{2\pi^2}\left|\frac{v_{T,k}(\eta)}{z_T(\eta)}\right|^2=\frac{\rho_c^2}{\rho_{pl}}R_T^2,
\end{eqnarray}
where $R_T=\int_{-\infty}^{\infty}\frac{d\bar{\eta}}{z_T^2(\bar{\eta})}$. Consequently
using (\ref{powerspectrum}) and (\ref{powertensor}), one deduces that the quotient $r=\frac{{\mathcal P}_{T}(k)}{{\mathcal P}_{\zeta}(k)}$ is equivalent to the formula
(\ref{ratio}).
\end{remark}

A more involved problem appears when one deals with cosmological perturbations in the framework of holonomy corrected LQC. 
As we have already pointed out in  the introduction, in the super-inflationary phase the speed of sound becomes imaginary implying Jeans instabilities, and then, the use of linear
perturbation equations is questionable in this approach. From our viewpoint, this fact, shows that one cannot have too much confidence in the results obtained using 
the perturbation equations comming from holonomy corrections, 
which is not the case of the theoretical results obtained  from the perturbation equations in  $F(T)$ coming from the model given by equation (\ref{teleparallelLQC}).

\subsection{Numerical results for the current model}
Performing the change of variable $\tilde{t}=\sqrt{\rho_c}t$ in equation (\ref{ratio})
we can see that $r$ does not depend on the value of the critical density. Then, to show the viability of the model one has to evaluate numerically the tensor/scalar ratio for all the orbits that satisfy the
constrain ${\mathcal P}_{\zeta}(k) \cong 2\times 10^{-9} $, and
to check if, for that set of orbits, there is a subset which satisfy either $r$ is smaller than  $0.11$ (the latest  {\it Planck's} data constrain) or
$r=0.20^{+0.07}_{-0.05}$  (last BICEP2 data).

Our numerical study shows that \cite{ha14}:
\begin{enumerate}
 \item In holonomy corrected LQC, the minimal value of  ${\mathcal P}_{\zeta}(k)$ is obtained for the orbit that at the bouncing time satisfies $\bar\varphi\cong -0.9870$, for that
 orbit we have obtained ${\mathcal P}_{\zeta}(k)\cong 23\times 10^{-3}\frac{\rho_c}{\rho_{pl}}$.
 \item In teleparallel LQC the orbit which gives the minimal value of the power spectrum satisfies $\bar\varphi\cong -0.9892$ and the value of the power spectrum is
 approximately the same as in holonomy corrected LQC
 ${\mathcal P}_{\zeta}(k)\cong 40\times 10^{-3}\frac{\rho_c}{\rho_{pl}}$.
\end{enumerate}

For these orbits, in order to match with the current result, in both theories, one has to choose $\rho_c\sim 10^{-7}\rho_{pl}$ which is
$2$ orders greater than the value needed using the analytical solution.

We have also calculated the ratio of tensor to scalar perturbations using formula (\ref{ratio}), and obtaining as a minimal value $r=0$ for the orbits that, at the bouncing
time, satisfy $\bar\varphi\cong -1.205$ and $\bar\varphi\cong 1.205$ .
On the other hand, its maximal value $r\cong 6.7187$ is attained by the solution (\ref{sol}) bouncing
at $\bar\varphi =0$. Then, since  the value of the tensor/scalar ratio in admissible solutions
 ranges continuously from the minimal
value $r=0$,
to the maximal value $r\cong 6.7187$, one can deduce that there is a set of solutions which matches correctly with BICEP2 data and another one with {\it Planck's} result.

Then,  the confidence interval $r =0.20^{+0.07}_{-0.05}$
derived from BICEP2 data is realized by solutions bouncing when
$\bar\varphi$  belongs in $[-1.162,-1.144]\cup[1.144,1.162]  $, and the bound $r\leq 0.11$ provided by {\it Planck's} experiment is realized by
solutions bouncing  when $\bar\varphi$ is in the interval  $[-1.205,-1.17] \cup [1.17,1.205] $.
Moreover, subtracting various dust models the tensor/scalar ratio in BICEP2 experiment could be shifted to $r=0.16^{+0.06}_{-0.05}$ with $r=0$ disfavored
at $5.9\sigma$. Then,  this confidence interval is realized by solutions bouncing when
$\bar\varphi \in [-1.17, -1.151]\cup[1.151,1.17] $.

On the other hand, in holonomy corrected LQC, numerical results show that the allowed
 orbits provide values of $r$ in the interval $[0,0.1114]$, matching only correctly with
{\it Planck's} constrain $r\leq 0.11$. In figure $4$, we have drawn the tensor/scalar ratio for teleparallel and holonomy corrected LQC.

Finally, we have checked numerically that the functions $z''/z$ and $z_T''/z_T$ are increasing in the contracting phase and decreasing in the expanding one, for teleparallel and
holonomy corrected LQC. This means that all our formulae are correct, that is, all modes that leave the Hubble radius at early time satisfy the long wavelength relation
$|c_s^2| k^2\ll |z''/z|$ and $|c_s^2| k^2\ll |z_T''/z_T|$ up to late times, and thus, we can safely disregard the Laplacian term in the M-S equations, giving validity to our approximation.

\begin{figure}[h]
\begin{center}
\includegraphics[scale=0.30]{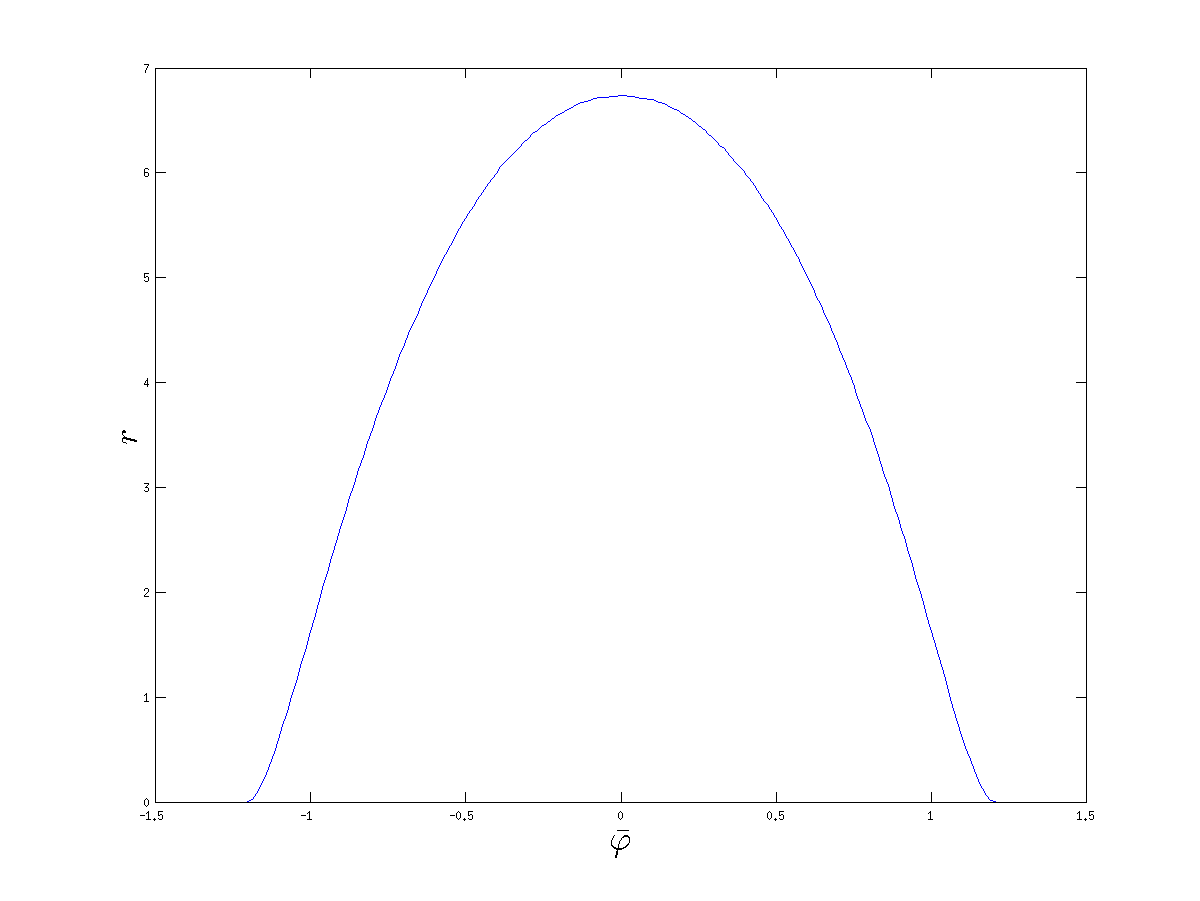}
\includegraphics[scale=0.30]{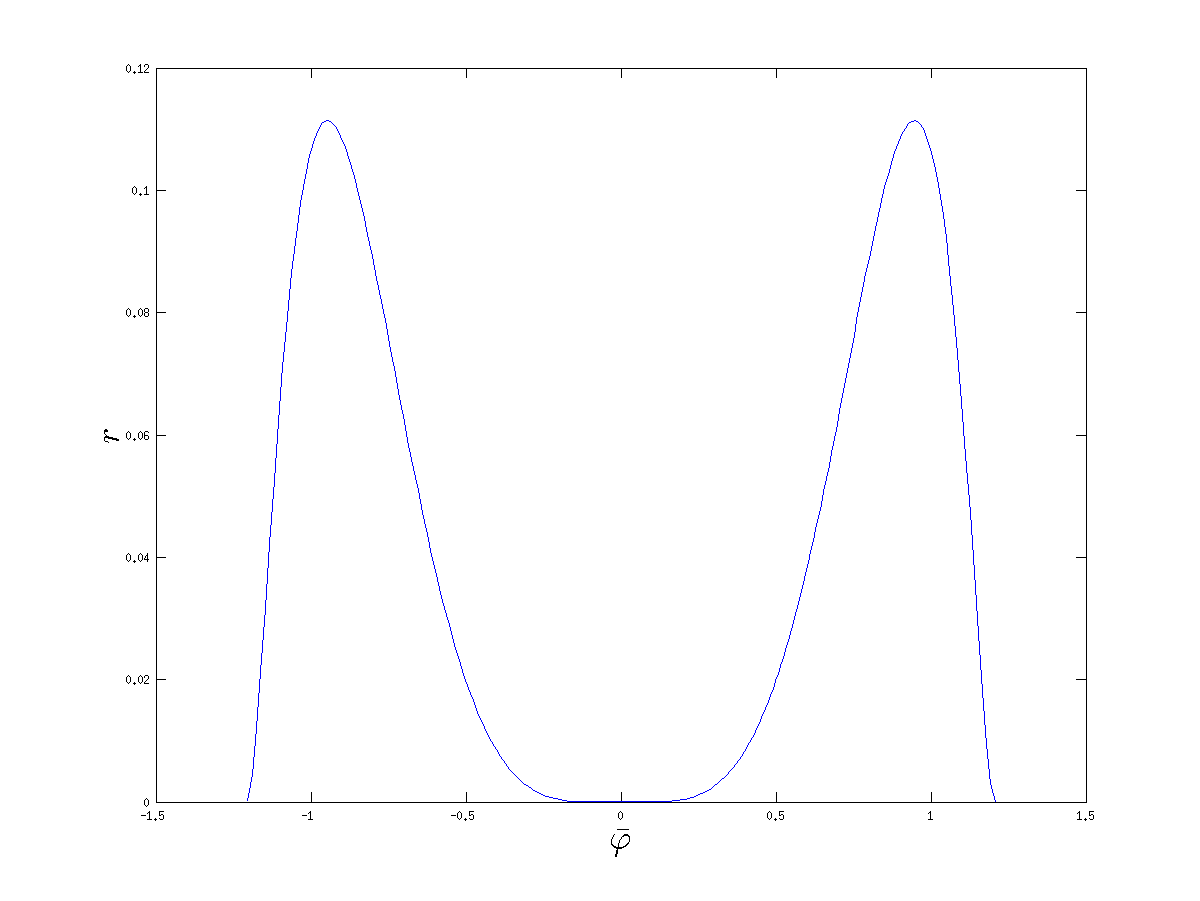}
\end{center}

\caption{{\protect\small Tensor/scalar ratio for different orbits  in function of the bouncing value of $\bar\varphi$. In the first picture for  teleparallel LQC, and in the second one for
holonomy corrected LQC.}}
\end{figure}

\section{Viable models for the matter bounce scenario}

According to the current observational data, in order to obtain a  viable  model in the matter bounce scenario in LQC, the bouncing model has
to satisfy  some conditions that we have summarized  as  follows:
\begin{enumerate}
 \item
 The latest Planck data constrain the value of the spectral index for scalar perturbations, namely $n_s\equiv 1+\frac{d \ln {\mathcal P}_{\zeta}(k) }{d \ln k}$,
to $0.9603\pm 0.0073$ \cite{Ade} (it is scale invariant with a slight red tilt).
It is well-known that the  ways to obtain a nearly scale invariant power spectrum of perturbations are either a quasi de Sitter phase in the expanding phase or a nearly
matter domination phase at early times, in the contracting phase \cite{w99}. Then, since
for the matter bounce scenario one has $n_s=1$,  if one wants to improve the model to match correctly with that data, one has to consider, at early times in the contracting phase,
a Universe with a equation of state $P=\omega\rho$. In that case, the spectral index is given by
$n_s=1+12\omega$ \cite{w13,h13}, and thus one has to choose $\omega=-0.0033\pm 0.0006$. Then,  the model for large values of the field
 ($\varphi\rightarrow -\infty$ or $\varphi\rightarrow \infty$ ) must satisfy
$V(\varphi)\sim \rho_c e^{-\sqrt{3(1+\omega)}|\varphi|}$, because for this kind of potentials  when $|\varphi|\rightarrow \infty$
all the orbits depict a Universe with equation of state $P=\omega \rho$ (this claim could be proved,  exactly in the same way,
as we have showed, at the beginning of Section $5$, the asymptotic behavior of this potential in the case $\omega=0$).

In this case the formula for the tensor/scalar ratio will become
\begin{eqnarray}
 r=\frac{1}{3(1+\omega)}\left(\frac{\int_{-\infty}^{\infty}\frac{1}{az_T^{2}}dt}{\int_{-\infty}^{\infty}\frac{1}{az^{2}}dt}\right)^2,
\end{eqnarray}
however, due to the small value of $\omega$, one can safely choose $\omega=0$ without changing significantly the results. In fact, numerically we have obtained that in teleparallel
LQC, when $\omega=-0.0033$ (which corresponds to $n_s=0.9603$) the ratio $r$ ranges continuously in the interval $[0,6.74]$, and thus there is a set of solutions satisfying BICEP2 data, 
and another one fitting well
with {\it Planck's} results. On the other hand, for holonomy corrected LQC when $\omega=-0.0033$, $r$ ranges continuously in the interval $[0, 0.10]$, meaning that
its theoretical results only match with {\it Planck's} data. 

 

Here an important remark is in order. One could argue that constraining the parameter $\omega$ to be $\omega=-0.0033\pm 0.0006$ to obtain a correct spectral index has to be considered as 
a fine-tuning. However,
in any inflationary scenario that involves a slow rolling scalar field, the same kind of fine-tuning must be done. To show that, we will consider
quasi-matter domination, i.e., $\dot{\varphi}^2\cong 2V\Longrightarrow\ddot{\varphi}\cong V_{\varphi}$, at very early times in the contracting phase (see \cite{eho} for all the details).
In this regime the Friedmann
and conservation equations  become
 \begin{eqnarray}\label{x2}
          H^2= \frac{2}{3}V,\quad
          3H\dot{\varphi}+2V_{\varphi}=0.
 \end{eqnarray}

 Introducing, in the same way as in slow roll inflation,  a {\it  quasi-matter domination parameter} 
\begin{eqnarray}\label{x3}
 \bar\epsilon
 =
 \frac{1}{3}\left(\frac{V_{\varphi}}{V} \right)^2-1,  
\end{eqnarray}
a simple calculation leads to
\begin{eqnarray}\label{x10}
 \frac{z''}{z}\cong\frac{1}{\eta^2}\left(\nu^2-\frac{1}{4} \right),
\end{eqnarray}
where
$
 \nu\cong \frac{3}{2}-6\bar\epsilon.
$
 And thus, the spectral index for scalar perturbations, namely $n_s^{MB}$,  is given by
 \begin{eqnarray}\label{x12}
 n_s^{MB}-1\equiv 3-2\nu=12\bar\epsilon.
\end{eqnarray}

It is instructive, to compare this result with the one obtained in inflationary cosmology
\begin{eqnarray}\label{x15}
 n_s^{SR}-1=2\bar\eta_{sr}-6\bar\epsilon_{sr},
\end{eqnarray}
where $\bar\epsilon_{sr}$ and $\bar\eta_{sr}$ are the well known slow-roll parameters, and we have denoted by $n_s^{SR}$ the spectral index in inflationary cosmology.
Moreover, in inflationary cosmology, the scalar/tensor ratio, namely $r^{SR}$, is related with the slow-roll parameter $\bar\epsilon_{sr}$, by the relation
$r^{SR}=16\bar\epsilon_{sr}$ (Note that in the matter bounce scenario the tensor/scalar ratio (\ref{ratio}) does not depend on the quasi-matter domination parameters). 

As an example,
we will choose the  potential $V(\varphi)=V_0 e^{-\sqrt{3(1+\omega)}|\varphi|}$  \cite{lm} which, has as Equation of State $P=\omega \rho$, and thus leads to a power law expansion.
An easy calculation will provide
\begin{eqnarray}\label{x13}
 n_s^{MB}-1=12\omega, n_s^{SR}-1=-{3}(1+\omega) \mbox{ and } r^{SR}=24(1+\omega).
\end{eqnarray}

Both theories, matter bounce scenario and inflation, have to match with the current data in order to be viable. In this case, from last {\it Planck's} data, the spectral index is
given by $n_s=0.9603\pm 0.0073$, which means that:
\begin{enumerate}
 \item  In the matter bounce scenario, for the potential $V(\varphi)=V_0 e^{-\sqrt{3(1+\omega)}|\varphi|},$  to match with experimental data, one will have to 
 choose $\omega=-0.0033\pm 0.0006$ (nearly matter domination).
 \item In power law  inflation, for the  same potential, to match with the experimental value of the spectral index one has
 to choose $\omega=-0.9867\pm 0.0024$ (quasi de Sitter phase). 
  Moreover, for this potential, since the tensor/scalar ratio is given by
$24(1+\omega)$,  to fit well with {\it Planck's} date one has to choose $\omega\leq -0.9954$, which is incompatible with $\omega=-0.9867\pm 0.0024$, and consequently,
{\it Planck's} data disregard this model.
On the other hand, to match the ratio of tensor to scalar perturbations with BICEP2 date, one has to choose $\omega\in [-0.9937,-09887]$ that together with the condition
$\omega=-0.9867\pm 0.0024$, restrict the value of the parameter $\omega$ to be $\omega=-0.9890^{+0.0001}_{-0.0003}$.
\end{enumerate}

The same will happen for the other models, for instance, in chaotic inflation if one considers the quartic potential $V(\varphi)=\lambda \varphi^4$, 
one has
\begin{eqnarray}\label{x17}
 n_s^{SR}-1=-\frac{24}{\varphi_N^2} \quad \mbox{and} \quad r^{SR}=\frac{128}{\varphi_N^2},
\end{eqnarray}
where  $\varphi_N$ is the value of the scalar field $N$ e-folds before the end  of the slow-roll phase.
Then,
in order to match with
BICEP2 data one has to choose $\varphi_N=22^{+0.5973}_{-0.2268}$, 
i.e.,
the theoretical value of the spectral index for modes that leave the Hubble radius approximately $N$ e-folds before the end of inflation, matches correctly with the experimental data. 
This is once again a fine tuning, because  
for this potential the value of the field $N$ e-folds before the end of the inflationary period must be tuned. In contrast, if one considers, in matter bounce scenario, a potential with the
asymptotic form
$V(\varphi)=V_0e^{\sqrt{3}\left(\varphi+\frac{1}{\varphi}\right)}$ {when}  $\varphi\ll 0$, one also obtains
$ n_s^{MB}-1=-\frac{24}{\varphi_N^2} $, where  $\varphi_N$ is the value of the scalar field, in the contracting phase, $N$ e-folds before the end  of the quasi-matter 
domination period.   Then,
in order to match with the experimental spectral index, one has to
choose $ \varphi_N\in [- 27.2165,-22.5973]$,
that is,
the theoretical value of the spectral index for modes that leave the Hubble radius approximately $N$ e-folds before the end of the quasi-matter domination epoch,
matches correctly with the experimental data.  

To sum up,
from our viewpoint, these calculations show that, in order to match the theoretical results with  current data, the parameters that appear in both theories
(inflation and matter bounce scenario), must be tuned finely.

\item The Universe has to reheat creating light particles that will thermalize matching with a hot Friedmann Universe. The simplest way to do that is with an oscillatory behavior of the
field in the expanding phase, because when the field oscillates it decays
releasing its energy at the bottom of the potential, where the adiabaticity of the process is strongly violated,  producing light particles \cite{kls}, whose number
 increases with each oscillation due to this broad parametric resonance regime.
To obtain this behavior, we can assume that in some region
 the potential has a minimum, i.e., it has a potential well. The simplest potentials with a minimum are the ones used in inflation, for example
 potentials with the same shape as the power law potentials used  in chaotic inflation, i.e.,
$V(\varphi)=\lambda\varphi^{2n}$.

Another way is the so called
{\it instant preheating}, where no oscillations are required \cite{fkl}. This mechanism works for potentials with a global minimum, but it is very efficient for potentials which slowly
decrease for large values of the scalar field as in the theory of quintessence \cite{fkla}, where, in the expanding phase, an inflationary potential is matched
with a quintessence one. Finally, reheating could also be produced due to the gravitational particle creation in an
expanding Universe \cite{geometric}. In this case, an abrupt phase transition (a non adiabatic transition) is needed in order to obtain sufficient particle creation that thermalizes producing a reheating temperature that fits well with current observations.
This method was used in the context of inflation in \cite{ford,peebles}, where a sudden phase transition from a quasi de Sitter phase
to a radiation domination or a quintessence phase was assumed in the expanding regime. We will show, at the end of this Section, that
gravitation particle production  could be applied to the matter bounce scenario, assuming a phase transition from the matter domination
to an ekpyrotic phase in the contracting regime, and obtaining a reheating temperature compatible with current data.

\item  Studies of distant type Ia
supernovae \cite{p99} (and others)  provide strong evidence that our
Universe is expanding in an accelerating way. A viable model must take into account this current acceleration, which could be incorporated, in the simplest case, with a cosmological
constant, or by quintessence models \cite{quintessence}.
Of course, there are other ways to implement the current cosmic acceleration, for example using $f(R)$ or $f(T)$ gravity, but the models
that provide this behavior are very complicated, and our main objective in this work is to present the simplest viable models.

\item The numerical  results (analytical ones will be impossible to obtain) calculated with the model have to match with experimental data, for instance, the power spectrum
of scalar perturbations has to be of de order
$10^{-9}$ and the ratio of tensor to scalar perturbations has to be either less than $0.11$ or in the range $r=0.20^{+0.07}_{-0.05}$, depending if one uses {\it Planck's} or
BICEP2 data.
The numerical calculations could be performed using formulae (\ref{powerspectrum}) and (\ref{ratio}),
where the quantities appearing in these formulae will be computed after solving numerically the conservation equation (\ref{KG}), where the scale factor, obtained numerically
integrating the equation $H=\frac{\dot{a}}{a}$, must satisfy at early times

\begin{eqnarray}
a(t)\cong
\left(\frac{3}{4}(1+\omega)^2\rho_ct^2\right)^{\frac{1}{3(1+\omega)}}\cong
\left(\frac{3}{4}\rho_ct^2\right)^{1/3}.\end{eqnarray}

Note that, if the potential is proportional to the critical density, performing the change of variable $\tilde{t}=\sqrt{\rho_c}t$, one can show that the tensor/scalar ratio is
 independent of $\rho_c$ and the power spectrum is proportional to $\rho_c$, which means that ${\mathcal P}_{\zeta}(k)=K\frac{\rho_c}{\rho_{pl}}$ where $K$ is a dimensionless quantity
independent of $\rho_c$, and thus, the experimental data  ${\mathcal P}_{\zeta}(k)\sim 10^{-9}$
is easily achieved choosing $\rho_c\sim \frac{10^{-9}}{K}\rho_{pl}$.

\item The model has to be stable, in the sense that, if an orbit depicting the Universe  satisfies all the previous requirements, then a small perturbation of this orbit also
has to satisfy them. Mathematically speaking, the set of orbits that satisfy all the requirements must have nonzero measure.
\end{enumerate}


First,
one could deal, for simplicity, with a quadratic potential
$V(\varphi)=\frac{1}{2}m^2\varphi^2$,
because at early  times (in the contracting phase)  the Universe is in a matter dominated phase (in fact, it is matter dominated on average over few oscillations, because when
$|t|\rightarrow \infty$
one has $H(t)\cong \frac{2}{3t}\left(1-\frac{\sin (2mt)}{2mt} \right)^{-1}$ \cite{m05}). Moreover, at late times in the expanding phase, the field oscillates around the minimum of its potential
releasing its energy and creating light particles, that finally thermalize yielding  a hot Friedmann Universe that matches with the Standard Model, but does not take into account
the current accelerated expansion of the Universe. Note that one cannot add to the model a simple cosmological constant $\Lambda$ because, in this case, the Universe would start, in
the contracting phase,
in an anti de Sitter stage which does not lead to a nearly scale invariant spectrum.

A more interesting model that takes into account the current cosmic acceleration is
obtained combining  $\rho_c e^{-\sqrt{3(1+\omega)}|\varphi|}$ with the quadratic potential and a small cosmological constant,
in the following way (subjected to the condition $V(\varphi)< \rho_c$),
\begin{eqnarray}
 V(\varphi)=\rho_c \frac{\frac{m^2\varphi^2}{2}+\Lambda}{\frac{m^2\varphi^2}{2}+\rho_0}\frac{1}{\cosh(\sqrt{3(1+\omega)}\varphi)},
\end{eqnarray}
where $\omega\approx -0.0033$,
 $\Lambda$ is a small cosmological constant and $\rho_0$ is an energy density parameter. Note that, for large values of $|\varphi|$ one has
$V(\varphi)\sim 2\rho_c e^{-\sqrt{3(1+\omega)}|\varphi|}$,
meaning that, for large values of $|\varphi|$ the Universe is nearly matter dominated. On the other hand, for small values of $|\varphi|$, one has
$V(\varphi)\sim \frac{\rho_c}{\rho_0}\left(\frac{1}{2}m^2\varphi^2+\Lambda\right)$, then for these values the field will oscillate in the well of the potential releasing its
energy and matching with the
$\Lambda$CDM model, or an instant preheating will occur, before the field starts to oscillate, to match with the  $\Lambda$CDM model.

From the model we could see that orbits starting at $|\varphi|=\infty$ and ending at $\varphi=0$, are the ones that,  at early times, are in the contracting nearly matter dominated
phase and, at late times, match with the $\Lambda$CDM model. These orbits are the candidates to describe a viable Universe, the other
ones do not accomplish some of the requirements established by observations. Some of these orbits  start and end at $|\varphi|=\infty$, giving a matter
dominated Universe at early and late times, which contradicts the current acceleration of the Universe.
The other ones start
at early times, in the contracting phase, at the bottom of the well, i.e., starting in an anti de Sitter stage
(these orbits do not give a nearly invariant spectrum of perturbations). There are two kinds of the latter orbits:
\begin{enumerate}
 \item  The ones that leave the potential well
and finish at late times
in the expanding phase, in a nearly matter dominated phase.
\item The ones that do not clear the potential well and finish, in the expanding phase, in a de Sitter stage driven by
the cosmological constant $\Lambda$.
\end{enumerate}

For the orbits that depict a candidate to be a viable Universe one has to compute  the ratio of tensor to scalar perturbations, which is given by
formula (\ref{ratio}),
and to check if $r$ is smaller than  $0.11$ (the last  {\it Planck} data) or it is in the range $r=0.20^{+0.07}_{-0.05}$ (the last BICEP2 data).

Note finally that, the value of the cosmological constant in Planck units is of the order $10^{-120}$ (see for instance \cite{bou}), which means that, when one makes numerical
calculations, its
value can be considered zero.

\vspace{0.25cm}

Another way to build models that satisfy the requirements would be to match a potential of the asymptotic form $V(\varphi)\sim \rho_c e^{-\sqrt{3(1+\omega)}{|\varphi|}}$
which leads to an nearly matter dominated phase at early times with an inflationary potential, for example:

\begin{enumerate}
 \item
Matching with a power law potential
\begin{eqnarray}\label{powerlaw}
 V(\varphi)=\kappa\rho_c \left\{\begin{array}{ccc}
\frac{e^{\sqrt{3(1+\omega)}{\varphi}}}{\left(1+e^{-\sqrt{3(1+\omega)}{\varphi}}\right)^2}& \mbox{for} &\varphi<\varphi_0\\
\lambda(\varphi-\varphi_1)^{2n}+\frac{\Lambda}{\rho_0}& \mbox{for} &\varphi\geq \varphi_0,
                   \end{array}\right.
\end{eqnarray}
where $0<\varphi_0<\varphi_1$,
and  the parameters $\lambda>0$  and $\kappa>0$ are dimensionless.

One has to impose the continuity of $V$ and its first derivative at $\varphi=\varphi_0$.

In this model the orbits that could be acceptable are the ones that start at $\varphi=-\infty$ and end at the bottom of the
potential well ($\varphi=\varphi_1$). For such orbits one has
to calculate the corresponding power spectrum and the ratio of tensor to scalar perturbations and choose those whose theoretical values match correctly with  observations.

\item
Matching with a {\it plateau} potential
\begin{eqnarray}
 V(\varphi)=\kappa\rho_c \left\{\begin{array}{ccc}
e^{\sqrt{3(1+\omega)}{\varphi}}& \mbox{for} &\varphi<\varphi_0\\
\lambda\left(1-\frac{\varphi^2}{\varphi_1^2}\right)^{2}+\frac{\Lambda}{\rho_0}& \mbox{for} &\varphi\geq \varphi_0,
                   \end{array}\right.
\end{eqnarray}
where   $\varphi_0<0<\varphi_1$,  with $\lambda>0$ and $\kappa>0$  dimensionless parameters.
\end{enumerate}

The last way to build models consists in matching a potential with the asymptotic form $V(\varphi)\sim \rho_c e^{-\sqrt{3(1+\omega)}{|\varphi|}}$ with a quintessence potential,
for example with  the potential $V(\phi)=\rho_c\frac{\varphi_0^4}{\varphi^4+\varphi_0^4}$ introduced by Peebles and Vilenkin in
 \cite{peebles}.

\vspace{0.5cm}

Here an important remark is in order: In the expanding phase, reheating occurs when holonomy effects are negligible. Then,
the M-S variable $z$ will be given by
\begin{eqnarray}z(t)=\frac{a(t)\sqrt{-2\dot{H}(t)}}{H(t)}=\frac{a(t)\sqrt{\dot{\bar\varphi}^2+\rho_{\chi}+P_{\chi}}}{H(t)},
\end{eqnarray}
where $\rho_{\chi}(t)$ and $P_{\chi}(t)$ are the energy density and pressure  of the produced light $\chi$-particles
when adiabaticity is strongly violated. If reheating occurs via broad parametric resonance (the potential has a minimum),
between $15$ and $25$ oscillations are needed to complete the reheating \cite{m05},
and during these oscillations it is nearly impossible to describe analytically this process, i.e., it is nearly impossible to
have a formula for the evolution of $\rho_{\chi}$ and $P_{\chi}$, and consequently, since  is impossible to know the value of $z(t)$ during
the reheating one cannot calculate the ratio of tensor to scalar perturbations. For this reason, we will
 assume that reheating could only be done
 instantaneously \cite{fkl,fkla} in potentials with a minimum.
To be more precise, in the case of potentials with a global minimum, assuming instant reheating, we will
use the formula $z(t)=\frac{a(t)\dot{\bar\varphi}(t)}{H(t)}$ up to when the field arrives at the
minimum of the potential where adiabaticity is violated
and particles are created giving, nearly instantaneously,  a radiation dominated Universe. Then,  after reaching the minimum, since the Universe is radiation dominated,  we will use the formula
$z(t)=\frac{a(t)\sqrt{-2\dot{H}(t)}}{H(t)}=2a(t)$.
In contrast, for potentials without a minimum (for example, $V\sim e^{-\sqrt{3}|\varphi|}$ matched with a quintessence potential), gravitational particle creation explains the reheating of the Universe. In the case of inflationary cosmology, gravitational reheating
 is produced via an abrupt transition, in the expanding phase, from a quasi de Sitter regime to a quintessence one \cite{peebles}, but
 as we will see, in a bouncing scenario gravitational reheating could also be produced before the bounce, i.e., in the contracting phase.


\subsection{Numerical results}
In figure $5$ we have depicted the potential (\ref{powerlaw}) with $n=1$, where the matching has been done imposing the continuity of the first derivative.

\begin{figure}[h]
\begin{center}
\includegraphics[scale=0.30]{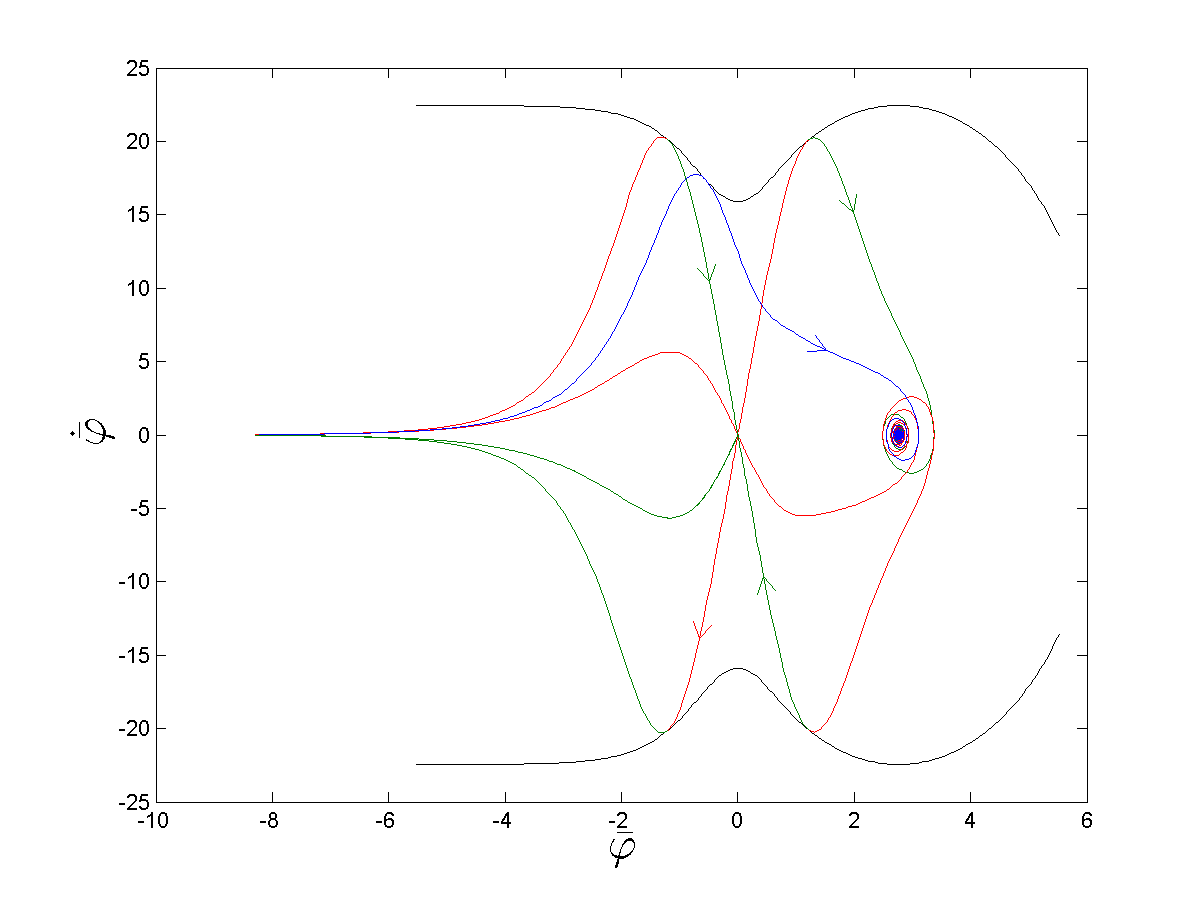}
\includegraphics[scale=0.30]{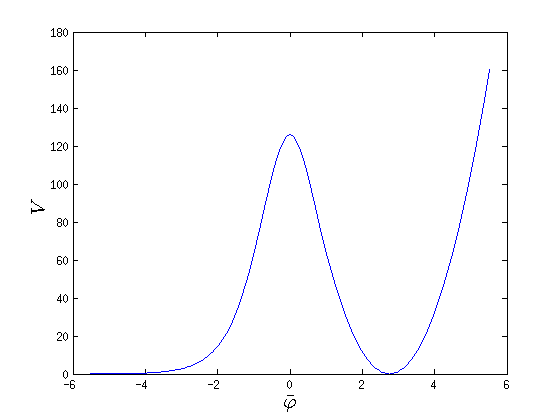}
\end{center}

\caption{{\protect\small In the first picture we have the phase portrait for the potential (\ref{powerlaw}) with $n=1$:  black curves   defined by $\rho=\rho_c$ depict the points where the Universe bounces. The point $(0,0)$ is a saddle point, red (resp. green) curves are the invariant
curves  in the contracting (resp. expanding) phase. The blue curve corresponds to an orbit of the system.
Note that  the allowed orbits are those that touch the
black curve in the region delimited by an unstable red curve and a stable green curve. Those orbits start in the contracting phase
depicting a matter dominated Universe and, when they touch the black curve, the Universe enters in the
expanding phase, oscillating around the minimum of the potential.
In the second picture we have drawn the shape of the matched quadratic ($n=1$) potential (\ref{powerlaw}).}}
\end{figure}

For this quadratic potential we have obtained the following results:
\begin{enumerate}\item
In teleparallel LQC,
the confidence interval $r =0.20^{+0.07}_{-0.05}$
derived from BICEP2 data is realized by solutions bouncing when
$\bar\varphi$  belongs in the interval $[-1.17,-1.145]\cup[1.107,1.14]$, and the bound $r\leq 0.11$ provided by {\it Planck's} experiment is realized by
solutions bouncing when $\bar\varphi$ is in the interval  $[-1.2039,-1.1685]\cup[1.1432,1.2081]$.
\item
In holonomy corrected LQC,
the confidence interval $r =0.20^{+0.07}_{-0.05}$
 is realized by solutions bouncing when
$\bar\varphi$  belongs in $[0.125,0.42]\cup[0.895,1.04]$, and the bound $r\leq 0.11$  is realized by
solutions
whose value at bouncing time is in the interval  $[-1.2039,0.1]\cup[1.05,1.2081]$. Note that, contrary to the simplest case, i.e., to the
current model provided by the potential (\ref{pot}), there are orbits whose theoretical results agree with BICEP2 data.
\end{enumerate}

The graphics of the tensor/scalar ratio, for teleparallel and holonomy corrected LQC, are depicted in figure $6$.

\begin{figure}[h]
\begin{center}
\includegraphics[scale=0.30]{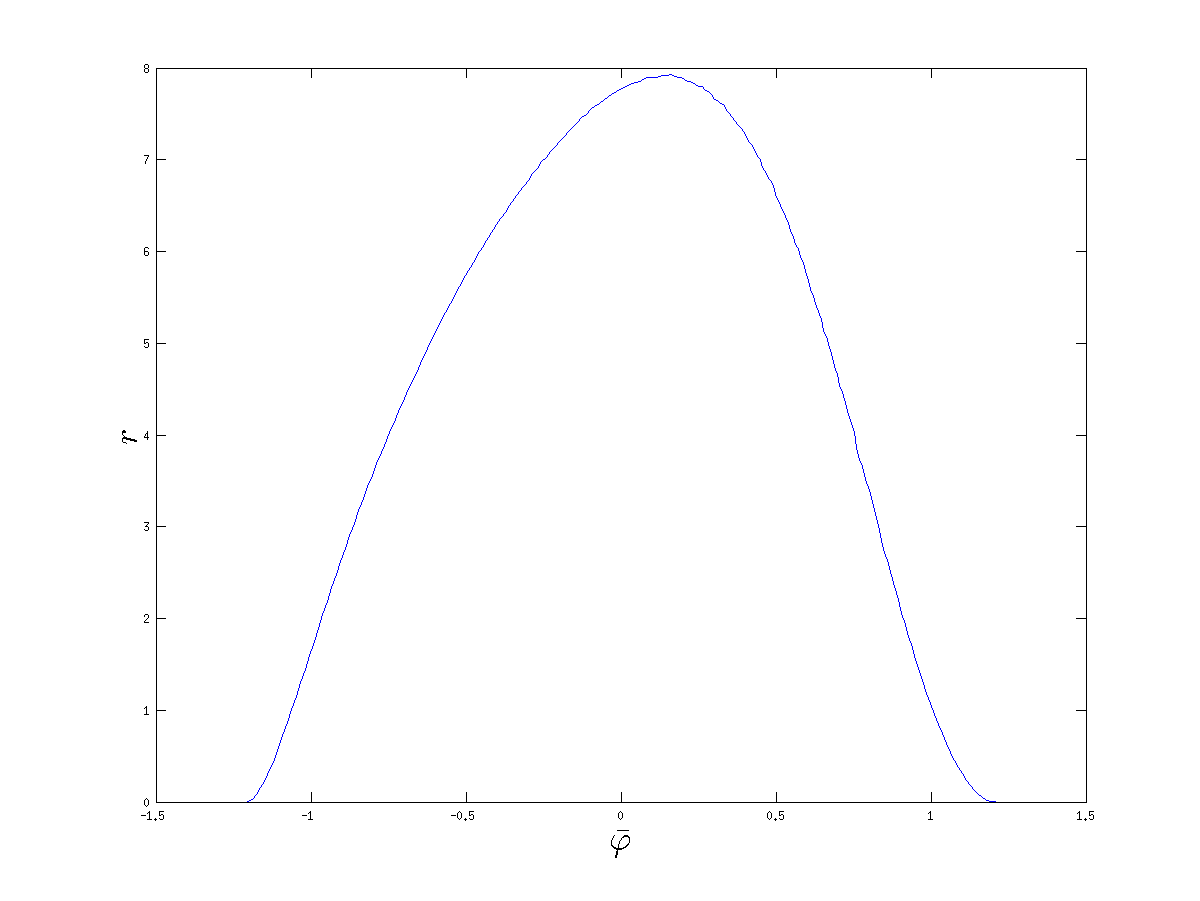}
\includegraphics[scale=0.30]{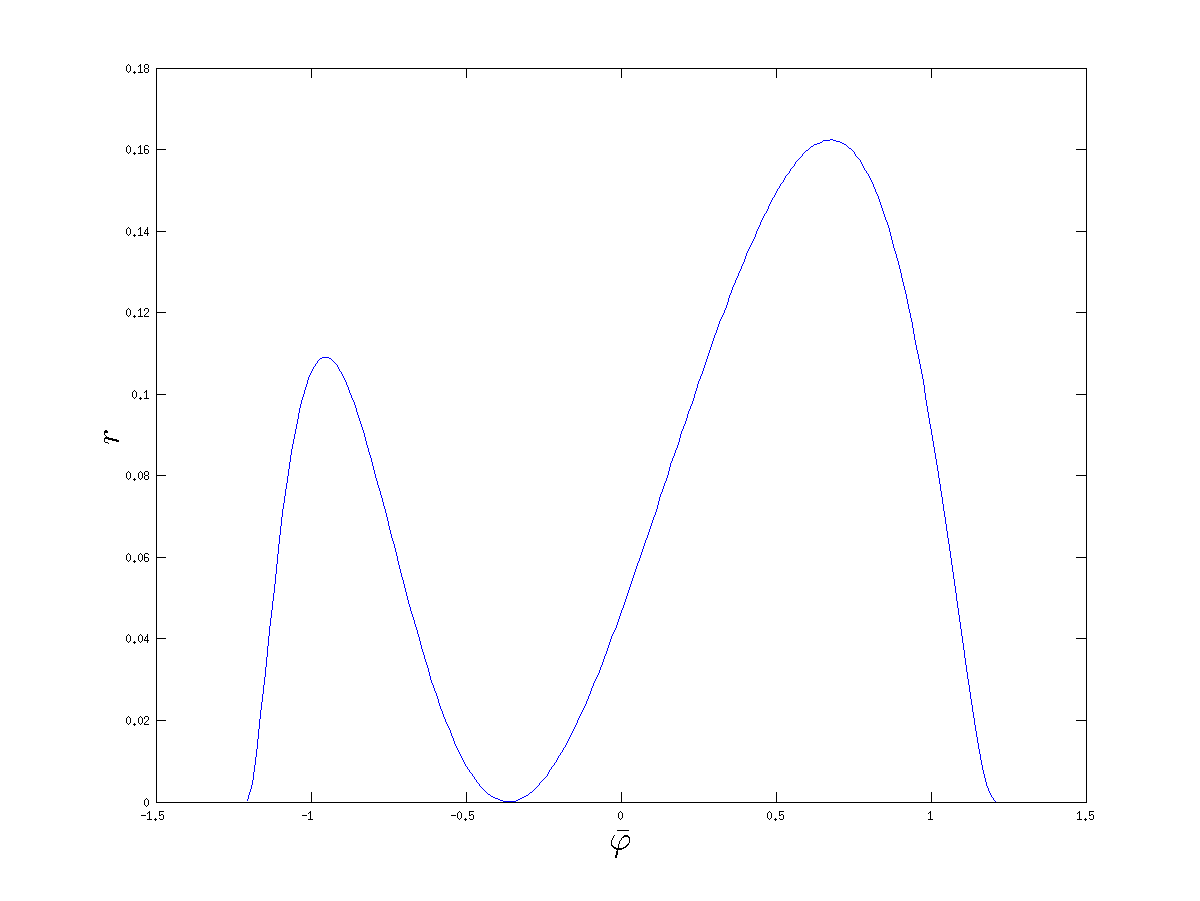}
\end{center}

\caption{{\protect\small Tensor/scalar ratio for different orbits  in function of the bouncing value of $\bar\varphi$ for the potential
(\ref{powerlaw}) with $n=1$. In the first picture for  teleparallel LQC, and in the second one for
holonomy corrected LQC.}}
\end{figure}

\vspace{1cm}

We have also studied numerically the potential (\ref{powerlaw}) for $n=2$ whose phase portrait and shape are given in figure $7$. For this model, we have
calculated numerically the corresponding ratio of tensor to scalar perturbations for different orbits, and  the results are depicted in figure $8$. From this last figure, we will see that in teleparallel LQC there are orbits whose theoretical results match correctly with BICEP2 data and others that fit well with last {\it Planck's} data. In contrast, in holonomy corrected LQC all the orbits provide a tensor/scalar ratio
smaller than $0.11$ what means that, for this model, holonomy corrected LQC only matches correctly with {\it Planck's} data.

\begin{figure}[h]
\begin{center}
\includegraphics[scale=0.30]{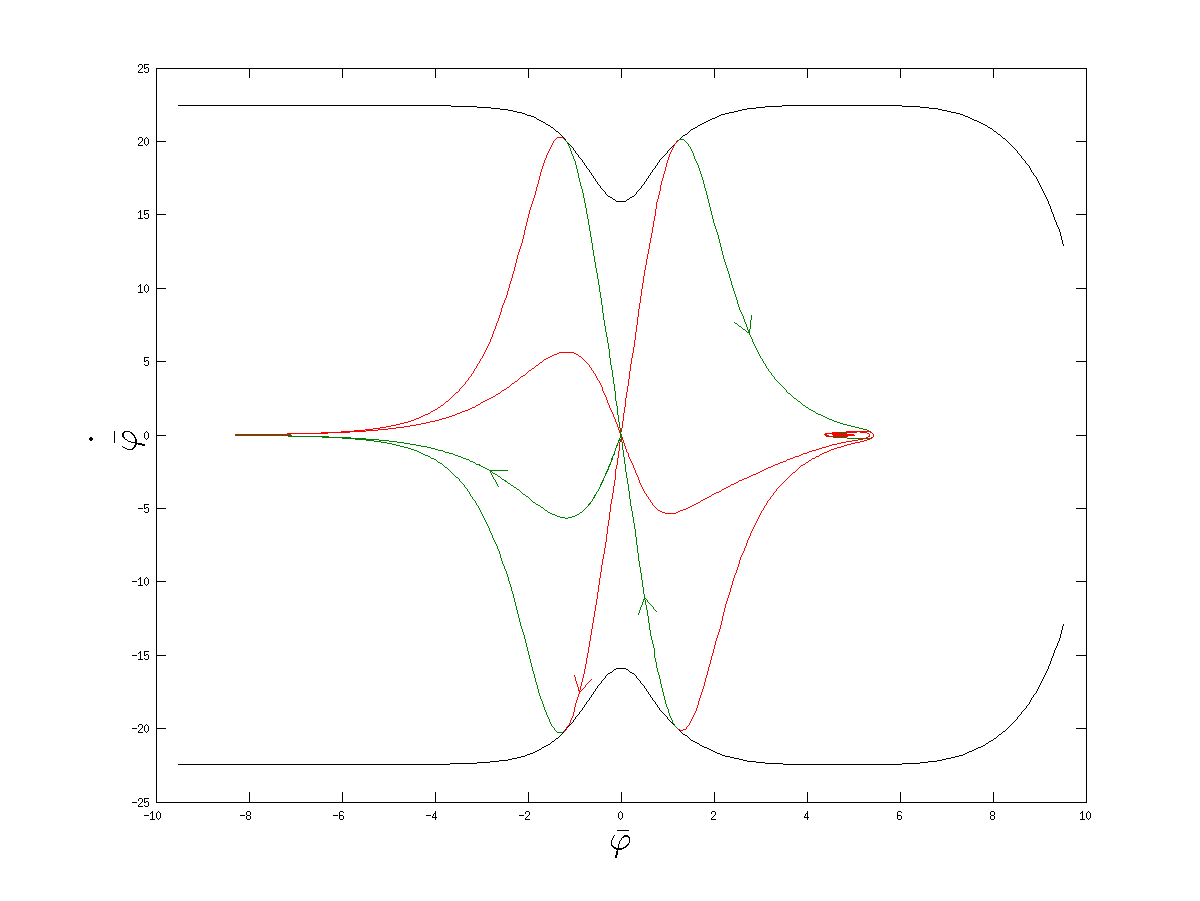}
\includegraphics[scale=0.30]{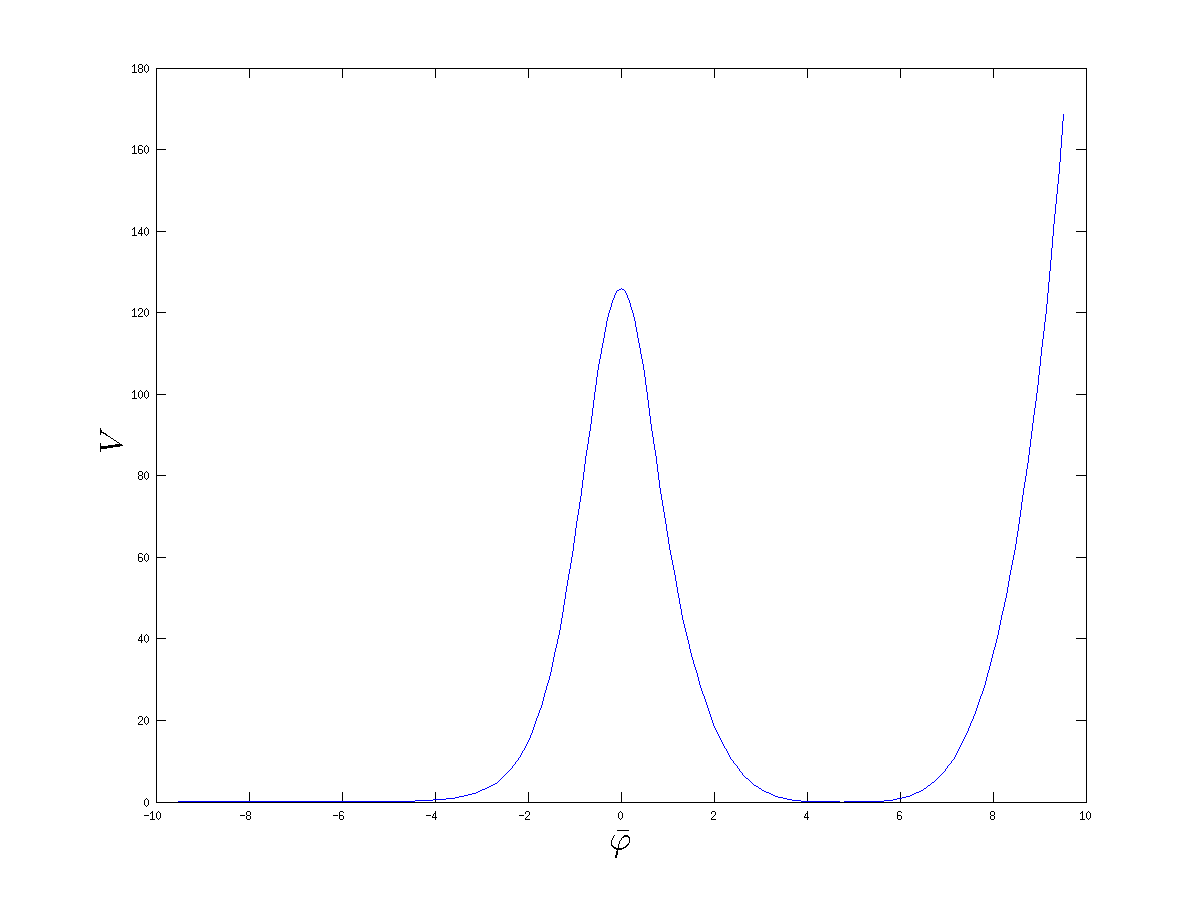}
\end{center}
\caption{{\protect\small Shape and phase space portrait of a potential that has matter domination at early times in the contracting phase and
and is matched with a quartic potential.}}
\end{figure}

\begin{figure}[h]
\begin{center}
\includegraphics[scale=0.30]{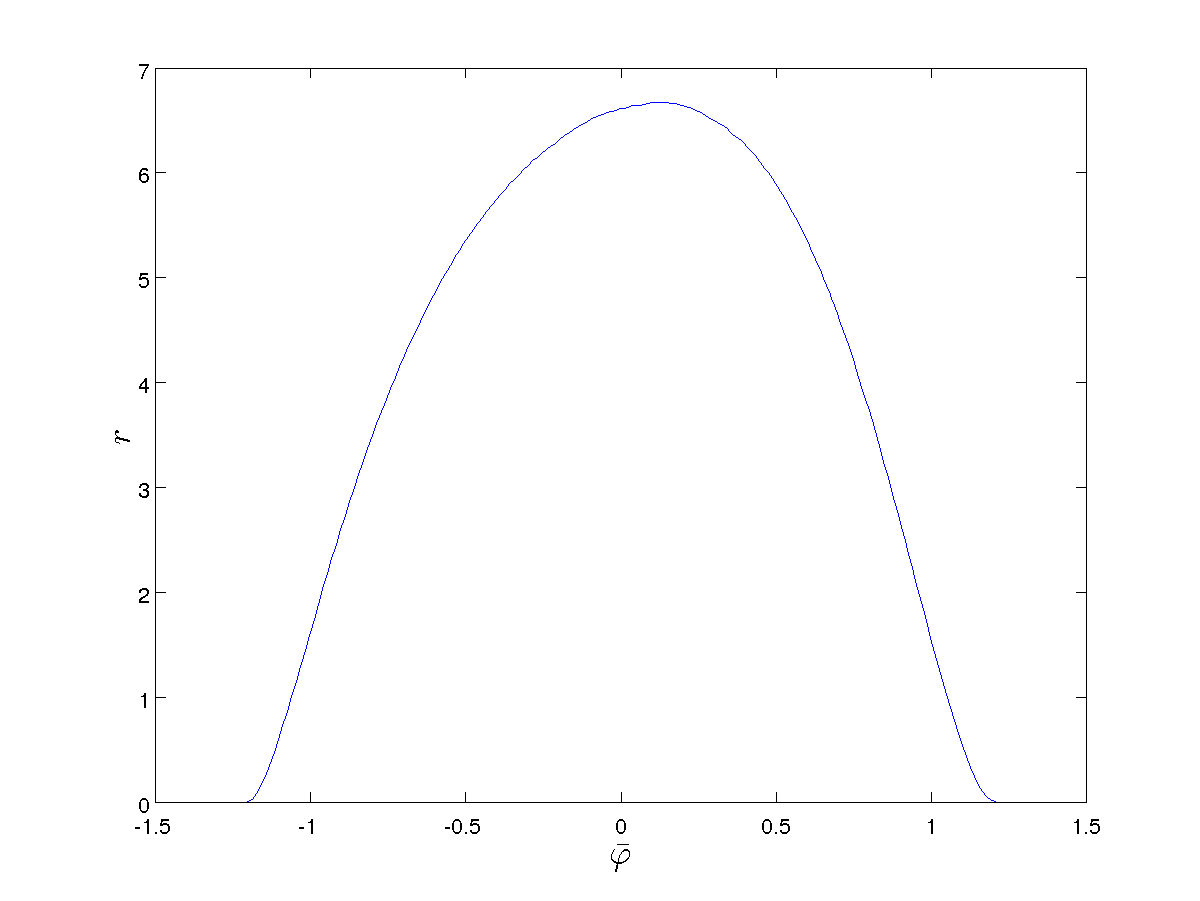}
\includegraphics[scale=0.30]{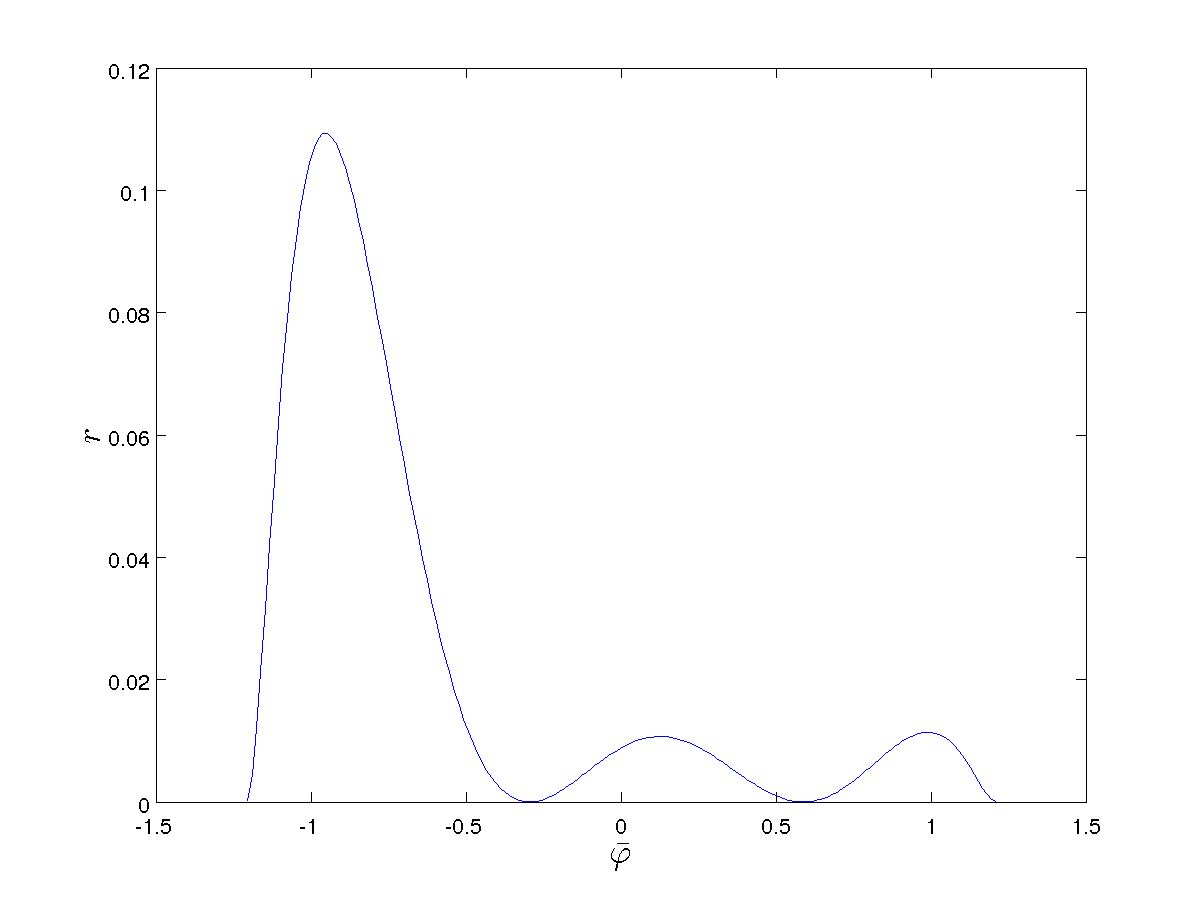}
\end{center}

\caption{{\protect\small Tensor/scalar ratio for different orbits  in function of the bouncing value of $\bar\varphi$ for the potential
(\ref{powerlaw}) with $n=2$. In the first picture for  teleparallel LQC, and in the second one for
holonomy corrected LQC.}}
\end{figure}

\vspace{1cm}

Finally, we have studied the model of quintessence matching, with continuous derivative,
a potential with the asymptotic form $V(\varphi)\sim \rho_c e^{-\sqrt{3(1+\omega)}{|\varphi|}}$ with a quintessence potential
for example, with  the potential $V(\varphi)=\rho_c\frac{\varphi_0^4}{\varphi^4+\varphi_0^4}$. In figure $9$, we show the shape of this potential and its phase portrait.

\begin{figure}[h]
\begin{center}
\includegraphics[scale=0.30]{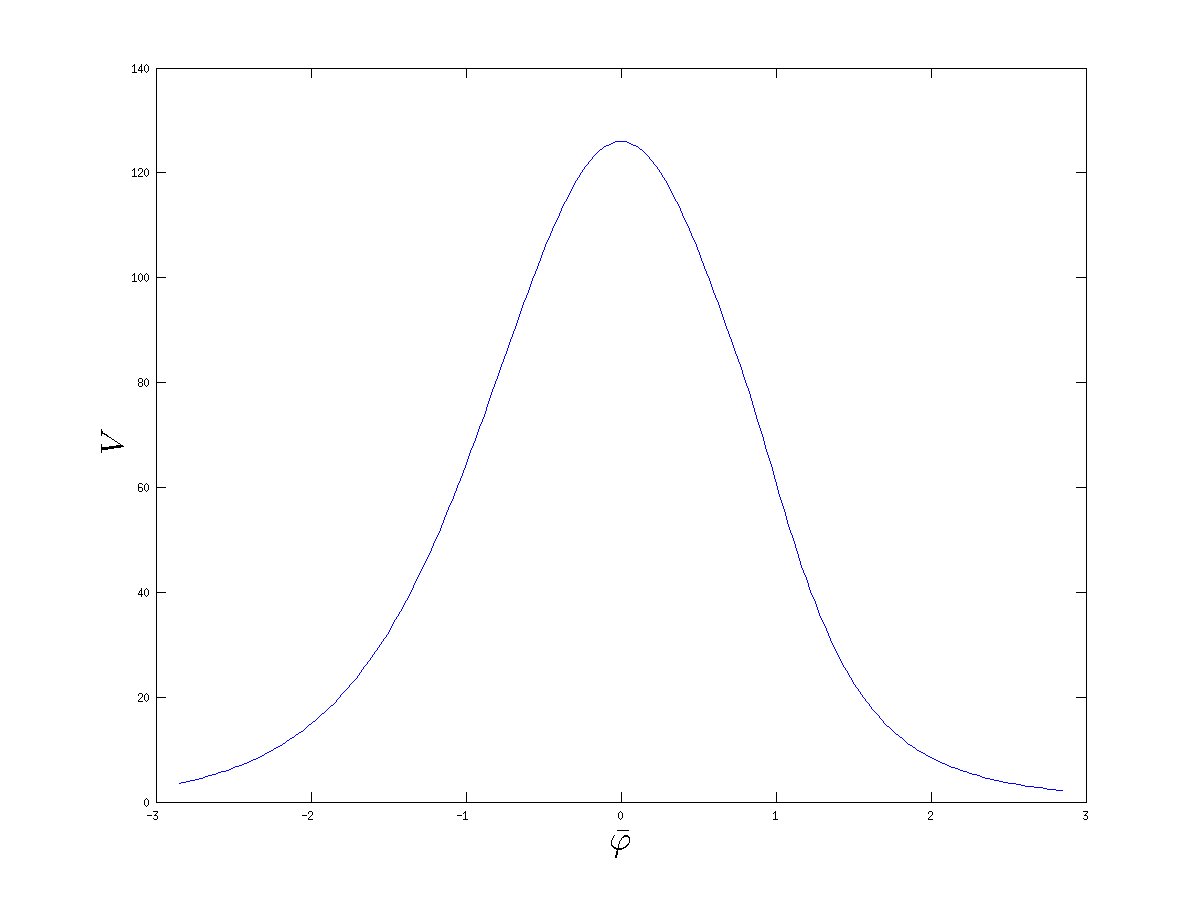}
\includegraphics[scale=0.30]{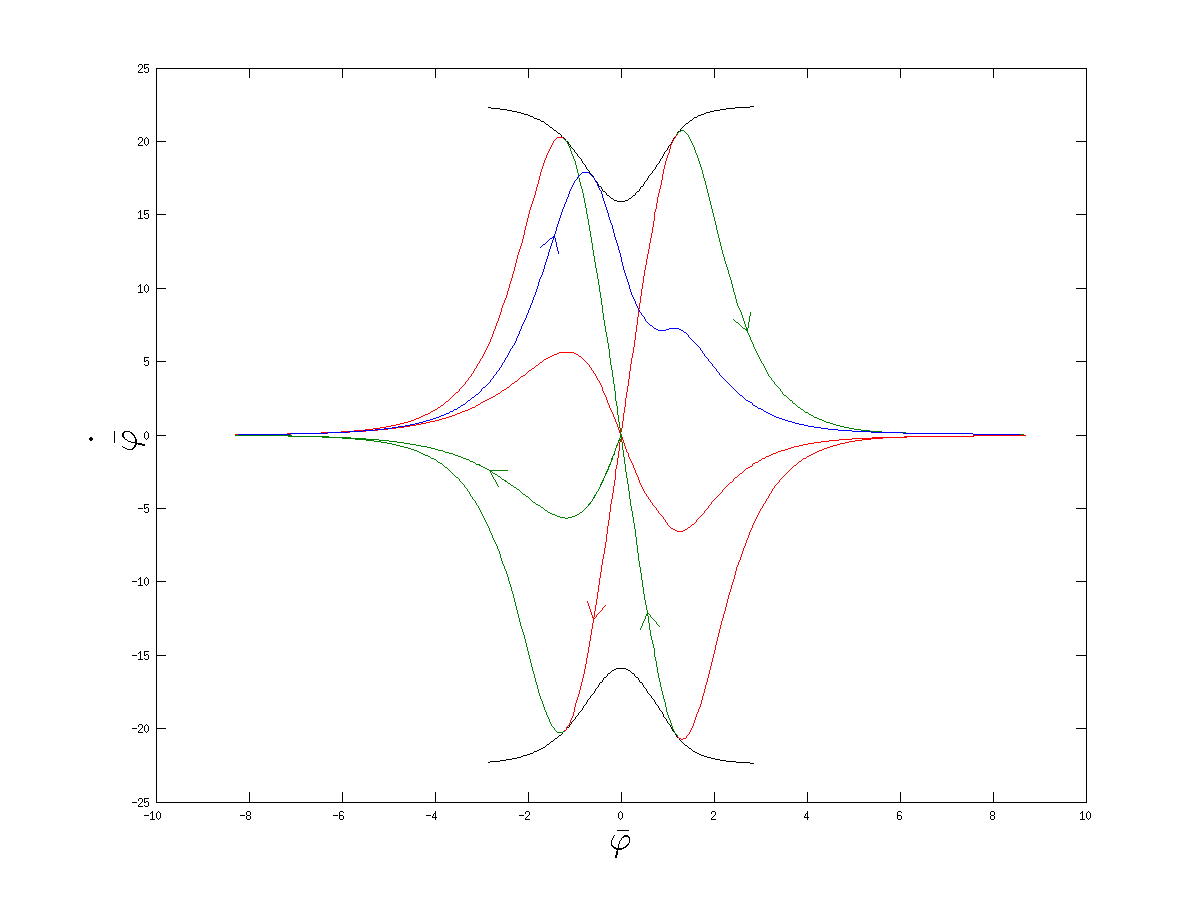}
\end{center}

\caption{{\protect\small Shape and phase space portrait of a potential that has matter domination at early times in the contracting phase and quintessence at late
times in the expanding phase.}}
\end{figure}

For this potential we have obtained the following results:
\begin{enumerate}\item
In teleparallel LQC,
the confidence interval $r =0.20^{+0.07}_{-0.05}$
derived from the BICEP2 data is realized by solutions bouncing when at bouncing time
$\bar\varphi$  belongs in the interval $[-1.162,-1.145]\cup[1.135,1.155] $, and the bound $r\leq 0.11$ provided by {\it Planck's} experiment is realized by
solutions whose value at bouncing time is in the interval  $[-1.208,-1.17]\cup[1.164,1.204] $. Moreover, if one considers BICEP2 subtracting various dust models the tensor/scalar
ratio, has we have already explained, is shifted to $r=0.16^{+0.06}_{-0.05}$. Then, theoretical results fit well when at bouncing time the value of the field
  belongs in $[-1.17,-1.152]\cup[1.143,1.164] $.
\item
In holonomy corrected LQC,
the confidence interval $r =0.20^{+0.07}_{-0.05}$
 is never realized because the maximum value of $r$ is $0.115$, and the bound $r\leq 0.11$  is realized by
solutions whose value at bouncing time belongs in the interval  $[-1.208,-0.995]\cup[-0.865,1.204]
$.
\end{enumerate}

The graphics of the tensor/scalar ratio are depicted in figure $10$

\begin{figure}[h]
\begin{center}
\includegraphics[scale=0.30]{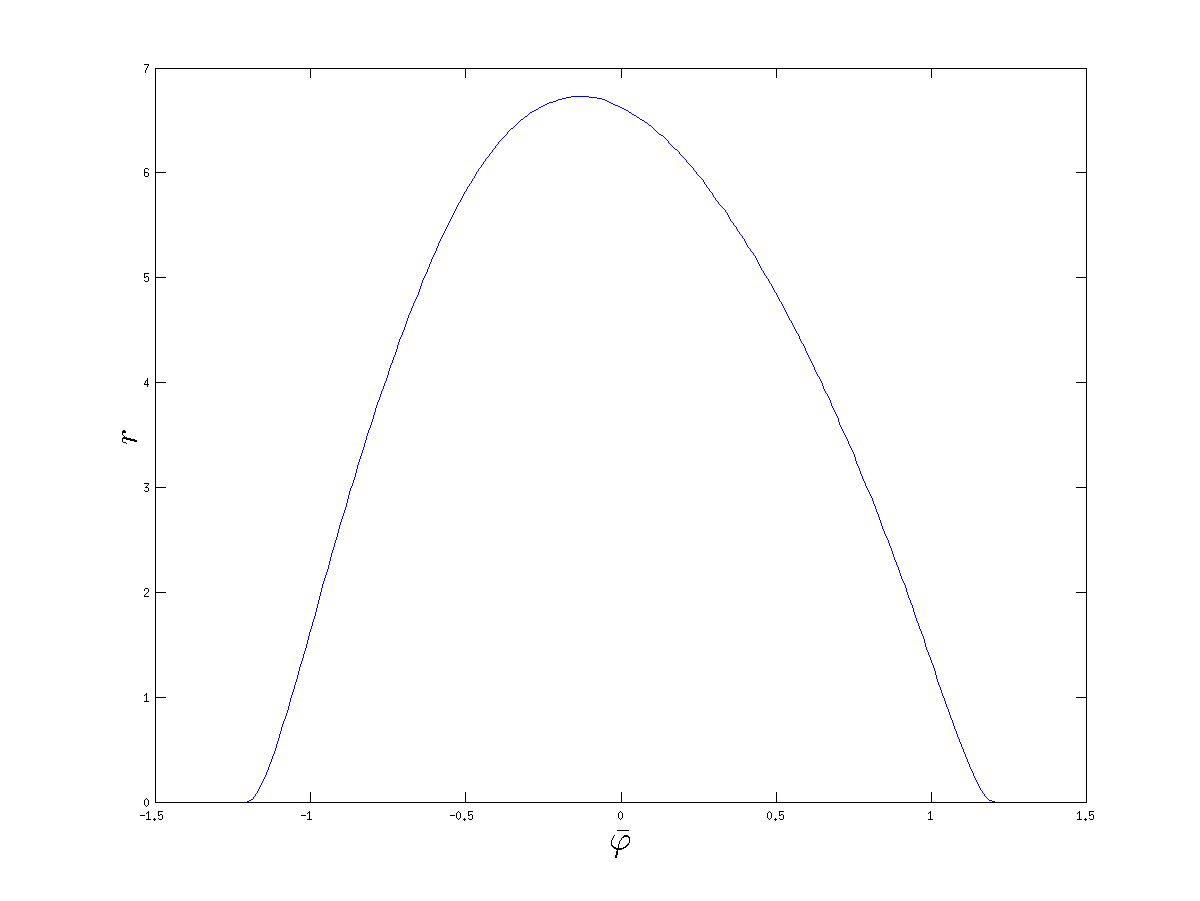}
\includegraphics[scale=0.30]{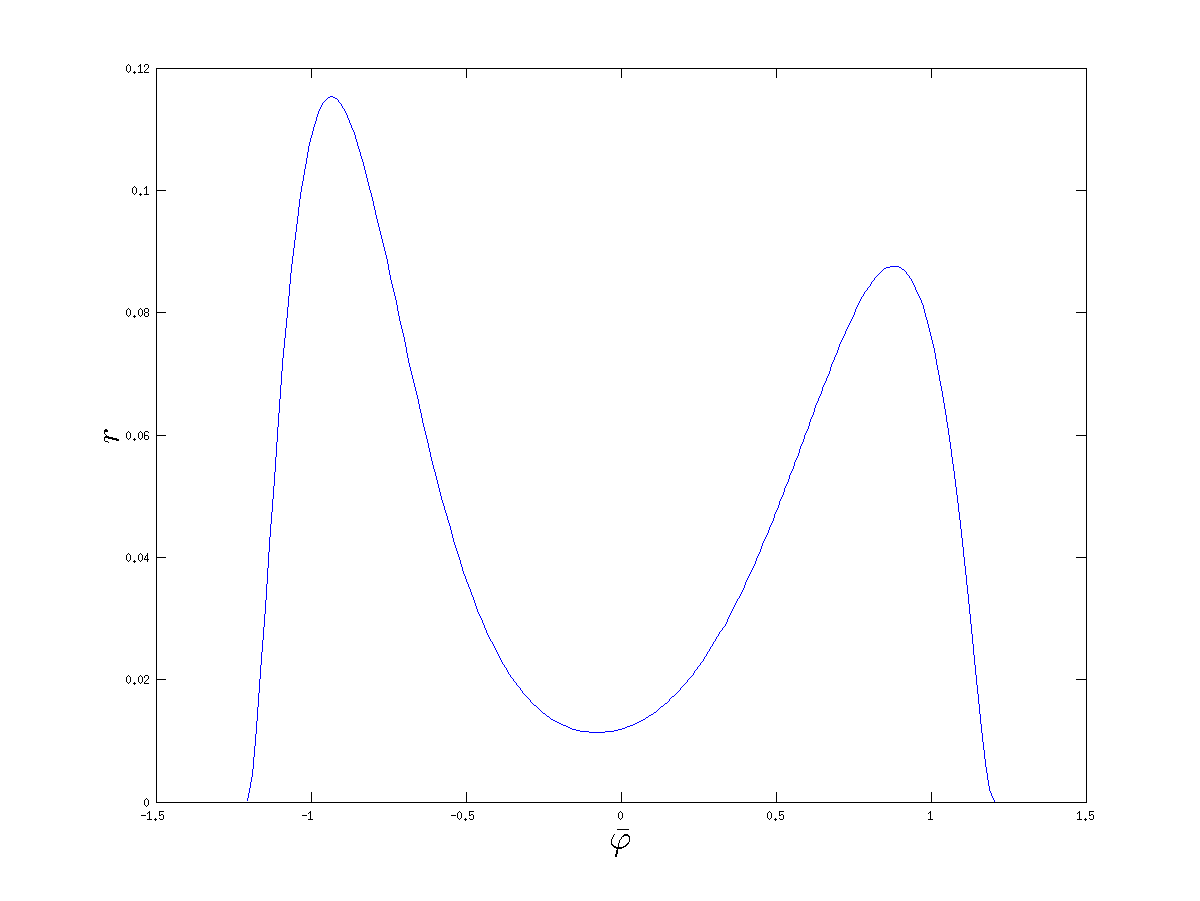}
\end{center}

\caption{{\protect\small Tensor/scalar ratio for different orbits  in function of the bouncing value of $\bar\varphi$ for the quintessence potential. In the first picture for
teleparallel LQC, and in the second one for
holonomy corrected LQC.}}
\end{figure}

As we will see from these numerical results, the shape of the tensor/scalar ratio in teleparallel LQC is very robust, in the sense that,
it is potential independent (is nearly the same for all the viable potentials we have studied). However, dealing with holonomy corrected
LQC, one can see that the shape of the ratio of tensor to scalar perturbations and the theoretical results obtained from it, change completely depending on the potential chosen.

\subsection{Reheating via gravitational particle production}
Gravitational particle production in the matter bounce scenario has been recently introduced in \cite{Quintin}. The idea is the same
as in inflationary models with potentials without a minimum (the so-called non oscillatory models): to have an efficient reheating one needs a non-adiabatic  transition, in the expanding regime, between two different phases in order to have
enough  gravitational particle creation. In inflation, there is an abrupt transition from a quasi de Sitter regime to a radiation-dominated one,
during this transition light particles are created and its density evolves like $\rho_r\sim a^{-4}$. On the other hand, after the end
of the quasi de Sitter phase, the inflaton
field, namely $\phi$, enters a kinetic-dominated period where the energy density of the inflaton field evolves like
$\rho_{\phi}\sim a^{-6}$ \cite{ford, peebles, spokoiny}, which means that the inflation energy density decreases faster than that of
radiation, and thus, the Universe becomes radiation dominated and matches with the hot Friedmann universe. Only at very late times, if
the inflation potential is matched with a quintessence one, the Universe enters in an accelerated regime from which it never recovers.

In the matter bounce scenario the non-adiabatic  transition could be produced in the contracting phase. In fact, a transition from matter-domination
to an ekpyrotic phase with equation of state $P=\omega\rho$ where $\omega>1$   could be assumed in the contracting regime. The obtained model
is called matter-ekpyrotic bounce scenario \cite{caiewing}, and since in the ekpyrotic phase the energy density of the field evolves like $\rho_{\varphi}\sim a^{-3(1+\omega)}$, which in  the contracting phase increases faster than  $a^{-6}$, anisotropies become negligible
(note that the energy density in the anisotropies grows in the contracting phase as $a^{-6}$ which is  faster than the energy density of radiation, and thus, without an ekpyrotic transition the isotropy of the bounce is destroyed;
this is the so-called Belinsky-Khalatnikov-Lifshitz instability \cite{lifshitz}).
 Moreover, the
energy density of the field also increases faster than that of radiation, this
 means that the field dominates the evolution of the Universe in the contracting phase, but when the Universe bounces its energy density
eventually dominates, because in the expanding phase $a^{-3(1+\omega)}$ decreases faster than $a^{-4}$. Then, the Universe
 will become radiation-dominated, and only at late times, if the potential is matched with an quintessence one,  the Universe will
 accelerate forever.

To be more specific, we will study reheating via massless $\chi$-particles nearly conformally coupled with gravity, using the method developed in
\cite{birrell}.  It is well known that the number density of created particles and their energy density is related via the
$\beta$-Bogoliubov coefficient as follows \cite{ford}
\begin{eqnarray}
n_{\chi}=\frac{1}{2\pi^2 a^3}\int_{0}^{\infty}|\beta_{k}|^2k^2dk,\quad \rho_{\chi}=\frac{1}{2\pi^2 a^4}\int_{0}^{\infty}|\beta_{k}|^2k^3dk,
\end{eqnarray}
where
\begin{eqnarray}
\beta_{k}=\frac{i(1-6\xi)}{2k}\int_{-\infty}^{\infty}e^{-2ik\eta}\frac{a''(\eta)}{a(\eta)}d\eta,\end{eqnarray}
being $\xi\cong \frac{1}{6}$ the coupling constant.

In this approach, the number density of produced particles could be easily calculated using the properties of the Fourier transform which lead
to the following simple formula
\begin{eqnarray}\label{number}
n_{\chi}=\frac{(1-6\xi)^2}{2\pi^2 a^3}\int_{-\infty}^{\infty}\left(\frac{a''(\eta)}{a(\eta)} \right)^2d\eta.
\end{eqnarray}

To perform the calculation we consider the simplest model of an abrupt transition from matter to ekpyrotic phase \cite{caiewing}
\begin{eqnarray}
a(t)=\left\{\begin{array}{ccc}
a_E\left(\frac{t-t_0}{t_E-t_0}\right)^{2/3}& \mbox{fot} & t\leq t_E\\
\left(\frac{3}{4}\rho_c(1+\omega)^2t^2+1\right)^{\frac{1}{3(1+\omega)}} &\mbox{for}& t\geq t_E,
\end{array}\right.
\end{eqnarray}
where $\omega\gg 1$, $t_0=t_E-\frac{2}{3H_E}$,  $t_E$ is the time at which the transition occurs and $H_E=H(t_E)$.
Note that  $t_0$ has been chosen in order that $a(t)$ has continuous first derivative at the transition time $t_E$.

For this scale factor, formula (\ref{number}) leads to following density of created particles
\begin{eqnarray}
n_{\chi}\cong \frac{3\sqrt{3}}{\pi^2}(1-6\xi)^2\omega\left(\frac{a_E}{a}\right)^3\rho_c^{3/2}\int_{-\infty}^{\infty}\frac{
(1-x^2)^2}
{(1+x^2)^{4}}dx,
\end{eqnarray}
and performing the integral one finally obtains
\begin{eqnarray}
n_{\chi}\cong \frac{3\sqrt{3}}{4\pi}(1-6\xi)^2\omega\left(\frac{a_E}{a}\right)^3\rho_c^{3/2},
\end{eqnarray}

The calculation of the energy density of the produced particles is more involved. First of all, to remove ultra-violet divergences one has to assume that the
second derivatives of the scale factor are continuous all the time. In this case, after integration by parts the $\beta$-Bogoliubov coefficient becomes
\begin{eqnarray}
\beta_{k}=-\frac{(1-6\xi)}{4k^2}\int_{-\infty}^{\infty}e^{-2ik\eta}\left(\frac{a''(\eta)}{a(\eta)}\right)'d\eta.
\end{eqnarray}

Now, for the sake of simplicity, we will assume as in \cite{ford} that the third derivative of the scale factor is discontinuous at the
transition time $\eta_E$. Then, one has
\begin{eqnarray}\label{f1}
\beta_{k}\cong\frac{(1-6\xi)}{8ik^3}e^{-2ik\eta_E}\frac{a'''(\eta_E^+)}{a(\eta_E)}\cong \frac{9(1-6\xi)}{16ik^3}e^{-2ik\eta_E}\omega^2H_E^3a_E^3,
\end{eqnarray}
where $a'''(\eta_E^+)$ is the value of the third derivative of the scale factor at the beginning of the ekpyrotic phase.

As a consequence, for modes well outside of the Hubble radius at the transition time, i.e., for modes satisfying
$a_EH_E<k<\infty$, using (\ref{f1})  one obtains the following radiated energy
\begin{eqnarray}
\rho_{\chi}\cong \frac{9}{16} (1-6\xi)^2\omega^3\frac{\rho_E^2}{\rho_{pl}}\left(\frac{a_E}{a}\right)^4,
\end{eqnarray}
where $\rho_E=3H_E^2$ is the energy density of the background at the transition time.

On the other hand, for modes well inside of the Hubble radius, i.e., satisfying $0<k<a_EH_E$, one can approximate the $\beta$-Bogoliubov coefficient by
\begin{eqnarray}
\beta_{k}=\frac{i(1-6\xi)}{2k}\int_{-\infty}^{\infty}\frac{a''(\eta)}{a(\eta)}d\eta
\cong -\frac{9i}{4} (1-6\xi)\frac{a_EH_E}{k},\end{eqnarray}
which means, because we have assumed $\omega\gg 1$, that the energy density of those modes is smaller  than that of the ones that are well outside of the Hubble radius, and consequently,
the total energy density of the produced particles is
\begin{eqnarray}
\rho_{\chi}\cong\frac{9}{16} (1-6\xi)^2\omega^3\frac{\rho_E^2}{\rho_{pl}}\left(\frac{a_E}{a}\right)^4
.\end{eqnarray}

To end the Section, we will calculate the reheating temperature $T_R$. To calculate that quantity  for our
model, first of all one has to define the reheating time $t_R$ as the time when the radiated energy density equals the background
energy density. Since the background energy in the ekpyrotic phase is given by
\begin{eqnarray}
\rho(t)=\frac{\rho_c}{\frac{3}{4}\rho_c(1+\omega)^2t^2+1},
\end{eqnarray}
for large values of $t$ one has $\rho(t)=\frac{4}{3\omega^2t^2}$. And thus, the reheating time is of the order
\begin{eqnarray}
t_R\sim \sqrt{\frac{\rho_{
pl}}{\rho_E}}\frac{1}{\omega^{5/2}|1-6\xi|H_E}.
\end{eqnarray}

Consequently, the reheating temperature is of the order
\begin{eqnarray}
T_R\sim \rho_r^{1/4}(t_R)\sim \sqrt{|1-6\xi|}\omega^{3/4}\frac{\rho_E^{1/2}}{\rho_{pl}^{1/4}}\frac{a_E}{a(t_R)}.
\end{eqnarray}

Finally,
writing $\rho_E$ in terms of Planck's density as follows $\rho_E\equiv \lambda^2 \rho_{pl}$, and approximating
$\frac{a_E}{a(t_R)} $ by $1$ because we are considering the case $\omega\gg 1$,
 one has
 \begin{eqnarray}
T_R\sim \rho_{\chi}^{1/4}(t_R)\sim \sqrt{|1-6\xi|}\omega^{3/4}\lambda M_{pl},
\end{eqnarray}
where we have introduced the Planck mass $M_{pl}\equiv \rho_{pl}^{1/4}$. This theoretical value  will coincide with current observations provided that one chooses $ \sqrt{|1-6\xi|}\omega^{3/4}\lambda\sim 10^{-7}$ \cite{Quintin}.

\begin{remark} Reheating via an interaction between the field $\varphi$ and light particles,
of the form $\epsilon^2\varphi^2\chi^2$,
has been recently studied in the context of bouncing scenarios in \cite{Quintin}. The calculation of the energy density
is very complicated, and only an upper bound has been obtained. Moreover, this energy density has been compared with the
energy density of produced   light particles minimally coupled with gravity, obtaining that the energy density of gravitational particle
production dominates over the energy density from the interaction for small values of $\epsilon$.
\end{remark}

\section{Conclusions} In this work we have shown that the matter bounce scenario in holonomy corrected LQC  and in its teleparallel version (the $F(T)$
model whose modified Friedmann equation coincides with the holonomy corrected one of LQC, when one considers the flat FLRW geometry) given by the
simplest potential (the potential given in  equation (\ref{pot})) must be improved in order to reproduce our
Universe. Although, when one considers our teleparallel version of LQC
the theoretical results predicted by some solutions of the potential (\ref{pot}) match  correctly either with BICEP2 or {\it Planck's} current data, this
potential
neither  provides a reheating mechanism  in order to match with a hot Friedman Universe nor
  takes into account the current acceleration of the Universe. For these reasons we
introduce more phenomenological potentials that take into account the
reheating process and the current cosmic acceleration, and we show that, in the teleparallel version of LQC, they have sets of solutions whose theoretical results fit well with current observational data,
concluding that
the matter bounce scenario in teleparallel LQC is a viable alternative to the inflationary paradigm. On the other hand, holonomy corrected LQC provides theoretical
results that always match with {\it Planck's} data, but in general, its theoretical results do not fit well with BICEP2 data, only for  some models  with a potential 
well (for example, for a quadratic potential there are solutions whose theoretical results match correctly with
BICEP2 data, but for a quartic one theoretical results only fit well with {\it Planck's} data).
In fact, our numerical results show that the teleparallel version of LQC is not potential dependent in our models, that is, we have shown that the shape
of the tensor/scalar ratio is practically the same for the models that we have studied. This does not happen in holonomy corrected LQC where
we have shown that it changes completely depending of the potential matched with (\ref{pot}).
From our viewpoint this is a great advantage of our $F(T)$ model with respect holonomy corrected LQC, but what really disfavours holonomy corrected LQC with respect our
teleparallel model, is that in this theory,
the speed of sound becomes imaginary in the super-inflationary phase, which implies Jeans instabilities, and these may invalidate the use of the linear perturbation equations in this regime. 
This problem does not arise in our $F(T)$ version of LQC, where the velocity of sound is always positive.

We  have also studied in detail the reheating process via gravitational particle production: we have considered the
gravitational reheating in the matter-ekpyrotic matter bounce scenario, where a phase transition from the matter domination to an ekpyrotic epoch occurs in the contracting phase 
of the Universe. We have assumed that the reheating is due to the creation of massless nearly conformally
coupled particles, and we have applied the method introduced by \cite{birrell}, for the first time, to our matter-ekpyrotic model, obtaining
a very simple expression for the reheating temperature.
Concluding that, when one considers potentials without a minimum,
the production of gravitational particles via a phase transition in the contracting regime
leads to an efficient, i.e. compatible with the latest experimental observations, reheating.
On the other hand, when one considers potentials with a minimum as in inflationary cosmology,
the instant reheating introduced in \cite{fkl} gives rise to a reheating temperature compatible
with current results.

Finally, we expect that more precise unified PLANCK-BICEP2 data (the B2P collaboration), which are going to be issued in the future, 
may be helpful to improve or rule out some of the bouncing models under consideration. 


\vspace{1cm}
{\bf Acknowledgments:}
This investigation has been
supported in part by MINECO (Spain), project MTM2011-27739-C04-01, MTM2012-38122-C03-01,
 and by AGAUR (Generalitat de Ca\-ta\-lu\-nya),
contracts 2009SGR 345, 994 and 1284.

\end{document}